\documentclass[amssymb,amsmath,12pt]{revtex4}
\usepackage{graphicx, color}
\usepackage{slashed}
\pdfoutput=1

\newcommand{\gsim}{\lower.7ex\hbox{$\;\stackrel{\textstyle>}{\sim}\;$}}
\newcommand{\lsim}{\lower.7ex\hbox{$\;\stackrel{\textstyle<}{\sim}\;$}}

\begin{document}

\baselineskip=16pt


\begin{center}

\bigskip
{\Large \bf 
 Signature of heavy sterile neutrinos at CEPC
}

\bigskip
{\bf Wei Liao and Xiao-Hong Wu}

\bigskip
{\it
Institute of Modern Physics,
School of Science, \\
East China University of Science and Technology,\\
Meilong Road 130,  Shanghai 200237, China
}
\end{center}

\centerline{\bf Abstract}
\begin{quote}
We study the production of  heavy sterile neutrino $N$, $e^+ e^- \to N \nu({\bar \nu})$,  at the 
Circular Electron Positron Collider(CEPC) and its $ljj$ signal in its decay to three charged fermions. 
We study background events for this process which are mainly events coming from W pair production. 
We study the production of a single heavy sterile neutrino and the sensitivity of CEPC to the mixing of 
sterile neutrino with active neutrinos. We study the production of two degenerate heavy sterile neutrinos 
in a low energy see-saw model by taking into account the constraints on mixings of sterile neutrinos from 
the neutrino-less double $\beta$ decay experiment and the masses and mixings of active neutrinos. We 
show that CEPC under proposal has a good sensitivity to the mixing of sterile neutrinos with active neutrinos 
for a mass of sterile neutrino around 100 GeV.

\end{quote}





\section{Introduction}

 The establishment of neutrino oscillation and tiny masses of active neutrinos in past decades has raised
 strong hope that new physics beyond the Standard Model(SM) is possible to exist in leptonic sector of elementary
 particles.  The see-saw mechanism \cite{seesaw}, as a simple and straightforward extension of neutrinos in the SM, works as a very good
 mechanism to explain the tiny masses of active neutrinos and is a very good candidate of physics beyond the SM.
 In see-saw mechanism, several right-handed neutrinos uncharged under the SM gauge groups, hence a type of sterile neutrinos,
 are introduced with heavy Majorana type masses which violate lepton number.  The tiny masses of active neutrinos are understood 
  in low energy scale as the lepton number violating remnant of the Majorana type masses of heavy right-handed neutrinos.
 
 Although see-saw type models are quite interesting models of physics beyond the SM and have fruitful implications,
 there are very few clues of the mass scale of  right-handed neutrinos. In particular, the mass scale of right-handed
 neutrinos can be much higher than the electroweak scale. Therefore, it is very hard to test such type models in experiments
 if such a hierarchy between the mass scale of right-handed neutrinos and the electroweak scale indeed exists. 
 For this reason, a low energy scale see-saw type model \cite{nuSM}, which has right-handed neutrinos at or below
 the electroweak scale, is quite interesting since it's possible to be tested in experiments.  There are several interesting
 properties of this low energy see-saw model. For example, one of the right-handed neutrinos can be of keV scale
 and serves as a good candidate of warm dark matter(WDM) in the universe. Two other right-handed neutrinos
 in the model
 are at GeV or hundred GeV scale and are sufficient to generate tiny masses and mixings of active neutrinos measured
 in neutrino oscillation experiments. 
 
 Another interesting property in this type of low energy see-saw model is that
 the Yukawa couplings of right-handed neutrinos with SM neutrinos can be quite large while they can still give
 rise to masses and mixing of active neutrinos being consistent with the experimental data in neutrino oscillation 
 and  the constraint from neutrino-less double $\beta$ ($0\nu\beta \beta$) experiment~\cite{Liao1, Liao2},
 in particular when two heavy right-handed neutrinos are degenerate or quasi-degenerate.
 Consequently,  the mixings of right-handed neutrinos with active neutrinos in the SM can be quite large while the
 masses of right-handed neutrinos are at GeV to hundred GeV scale.  This scenario apparently offers great opportunities
 to search for see-saw type of models of physics beyond the SM in collider experiment. 
 
Experimentally, the single heavy neutrino has been searched for
by L3 collaboration at LEP
through $N \to e W$ channel ~\cite{Acciarri:1999qj,Achard:2001qv}.
Stringent constraint on $|R_{eN}|^2$ has been set for a  mass region
from $80 {\rm GeV}$ to $205 {\rm GeV}$.
Some efforts have been made to study the
production and signature of heavy neutrino in $e^+ e^-$ or $e^- e^-$ collision processes
with both pair and single heavy neutrino productions,
and various neutrino decay chains, $lW$, $\nu Z$ and $\nu H$
~\cite{delAguila:1987nn,Buchmuller:1991tu,Gluza:1993gf,Djouadi:1993pe,
Azuelos:1993qu,Vuopionpera:1994kr,Gluza:1995js,Gluza:1996bz,Hofer:1996cs,
Gluza:1997ts,Cvetic:1998vg,delAguila:2005ssc,Das:2012ze,
Blondel:2014bra,Banerjee:2015gca,Caputo:2016ojx,
Antusch:2016vyf,Antusch:2016ejd,Biswal:2017nfl,Yue:2017mmi},
for a review, see~\cite{Antusch:2016ejd}.
Currently, new electron-positron colliders, 
such as CEPC, Future Circular Collider(FCC) and International Linear Collider(ILC), 
are under proposal. With these colliders,  heavy sterile neutrino can be probed to a larger  mass range
and  with better sensitivity on the active-sterile mixing $R_{lN}$.
Recently, single heavy neutrino production modes $N\nu$ and $Ne^\pm W^\mp$ at ILC with center of mass energy of
$350 {\rm GeV}$ and $500 {\rm GeV}$ have been investigated in~\cite{Banerjee:2015gca}.
A search of long-lived heavy neutrinos with displaced vertices
at CEPC, FCC and ILC has been presented in ref.~\cite{Antusch:2016vyf}.
In our work, we present a detailed study of $e^+e^- \to N\nu$
with charged current neutrino decay mode $N\to lW$
at CEPC with center of mass energy $\sqrt{s}=240 {\rm GeV}$.

In the present article, we are motivated by such kind of possibility and are going to
 study the signature of this type of right-handed neutrino( or to say sterile neutrino) of hundred GeV masses at 
 CEPC~\cite{CEPC}, a collider under proposal.  In the next section, we will make
 a quick review of the low energy see-saw model and describe some basic properties of this model. 
 Then we discuss the collider signatures of a single sterile neutrino of a mass around hundred GeV.
 For simplicity, we simplify our discussion of collider signature using a single sterile neutrino. 
 We will show that this simplification can be taken as a good simplification for later discussion.
 Then we come to signatures of low energy see-saw model by including detailed constraints on the masses
 and mixings of right-handed neutrinos. We conclude in the last section.  

\section{GeV scale sterile neutrino and low energy see-saw model }
One of major differences between the case of a single GeV scale sterile neutrino and the low
energy see-saw type model of GeV scale sterile neutrinos is that for the former the mixings
of sterile neutrinos with active neutrino are strongly constrained by  $0\nu\beta \beta$ decay experiment~\cite{0nubb},
while for the latter the $0\nu\beta \beta$ constraint can be quite weak and the mixings can be quite large~\cite{Liao1}.

In the presence of one or several sterile neutrinos, active neutrinos in the flavor base 
$\nu_l (l=e, \mu, \tau)$ are a mixture of  the light neutrinos in mass eigenstates $\nu_i (i=1, 2, 3)$
and heavy sterile neutrinos in mass eigenstates $N_j$,
\begin{eqnarray}
\nu_l = \sum_i U_{li} \nu_i + \sum_j R_{lN_j} N_j,
\end{eqnarray}
where $U_{lj}$ is the Pontecorvo-Maki-Nakagawa-Sakata(PMNS) mxing matrix,
and $R_{lN_j}$ is the matrix element mixing $\nu_l$ with heavy neutrinos $N_j$. 
For small enough $|R_{lN_j}|$,  mixing matrix $U$ can be considered as approximately unitary.
Apparently, $\nu_i$ and $N_j$
can all contribute, in virtual intermediate state, to the $0\nu\beta\beta$ decay.
It is not hard to see that the contribution of a single GeV scale sterile neutrino to the amplitude
of $0\nu\beta\beta$ decay is proportional to $R^2_{eN}/M_N$. 
The mixing $R_{eN}$  in this case is 
constrained to be $|R_{eN}|^2 \lsim 10^{-5}$~\cite{0nubb},  unless 
there are other particles or mechanisms at hand to ease the constraint. 

In low energy see-saw type model, at least two heavy sterile neutrinos(right-handed neutrinos)
are needed to obtain the correct masses and flavor mixings of active neutrinos~\cite{Liao1}.
In this case, the mixing matrix $R$ is $R=Yv (M^*)^{-1}$  where $Y$ is the Yukawa coupling
of neutrinos, v the vacuum expectation value in the SM and $M$ the Majorana mass matrix of sterile neutrinos
which can be taken to be real and diagonal in a convenient base.  The matrix M is a $2\times 2$ matrix if
considering two heavy sterile neutrinos and a $3\times 3$ matrix if considering three heavy sterile neutrinos.

A nice feature in see-saw model is that mixing $R$ is related to $m_\nu$,  the mass matrix of active neutrinos
responsible for the neutrino oscillation phenomena:
\begin{eqnarray}
(m_\nu)_{ll'}=-v^2 \sum_i Y^*_{li} Y^*_{l' i} M_i^{-1}=-\sum_iM_i R^*_{l N_i} R^*_{l' N_i}, \label{massrelation0}
\end{eqnarray}
where $M_i$ is the eigenvalue of matrix $M$,  that is, we have chosen a base in which $M$ is diagonal.
One can see that if a strong cancellation happens between contributions of different sterile neutrinos in 
(\ref{massrelation0}),  a mass matrix $m_\nu$ at $10^{-3}-10^{-2}$ eV scale can be generated for $M_i$ of
hundred GeV scale and for pretty large $|R_{lN_i}|$. 

Using mixing matrix $R$, contributions of heavy sterile neutrinos to the amplitude of $0\nu \beta\beta$ decay
can be parametrized as follows
\begin{eqnarray}
{\cal A}=F\sum_i R^2_{eN_i} M_i^{-1}, \label{0nubb-1}
\end{eqnarray}
where $F$ is an overall factor.  For two heavy sterile neutrinos $N_1$ and $N_2$,  (\ref{0nubb-1}) can be rewritten as
\begin{eqnarray}
{\cal A}=\frac{F}{M_1^2}( R^2_{eN_1} M_1+R^2_{eN_2} M_2 )
+F M_2 R_{eN_2}^2(\frac{1}{M_2^2}-\frac{1}{M_1^2}),\label{0nubb-2}
\end{eqnarray}
where $M_1$ and $M_2$ are the masses of $N_1$ and $N_2$ respectively. 
By taking $M_{1,2}$ real in a convenient base,
one can see in (\ref{massrelation0}) and (\ref{0nubb-2}) 
that the first term in (\ref{0nubb-2}) is of order $10^{-3}-10^{-2}$eV$/M_1^2$ and can be neglected.
The second term in (\ref{0nubb-2}) can be arbitrarily small if $N_1$ and $N_2$ are quasi-degenerate
or degenerate. One can see clearly that the constraint from $0\nu \beta\beta$ decay is no longer
strong for two quasi-degenerate heavy sterile neutrinos, which is exactly what happens in low energy
see-saw model.

A straightforward consequence of the above discussion about (\ref{0nubb-2}) and (\ref{massrelation0}) 
and the degeneracy of $N_1$ and $N_2$
is that for sterile neutrinos of GeV to hundred GeV mass, large value of $|R_{e N_i}|^2$ is only possible when 
\begin{eqnarray}
R^2_{eN_1}=  -R^2_{e N_2}, ~\textrm{ or}  ~R_{eN_1}=  \pm i R_{e N_2}. \label{massrelation1}
\end{eqnarray}
(\ref{massrelation1}) is one of the major relations to be used in later analysis for discussing the collider signal of
low energy see-saw model.  

Relations among $R_{\mu N_i}$ and $R_{\tau N_i}$ can also be addressed similarly. Using solutions presented
for two heavy sterile neutrinos in \cite{Liao1}, one can find that $R_{lN_i}$ can be expressed as
\begin{eqnarray}
R_{l N_1} =\frac{1}{2} e^{\mp i x+|y|}( U_{l2} m^{1/2}_2
 e^{-i \phi_2/2}\mp i U_{l3} m^{1/2}_3 e^{-i\phi_3/2}) (M^*_1)^{-1/2},  
 ~R_{l N_2}=\pm i R_{l N_1},\label{mixing1} 
 \end{eqnarray}
for normal hierarchy(NH) of neutrino masses, and
\begin{eqnarray}
 R_{l N_1} =\frac{1}{2} e^{\mp i x+|y|}( U_{l1} m^{1/2}_1
 e^{-i \phi_1/2}\mp i U_{l2} m^{1/2}_2 e^{-i\phi_2/2}) (M^*_1)^{-1/2},
 ~R_{l N_2}=\pm i R_{l N_1},  \label{mixing2} 
 \end{eqnarray}
for inverted hierarchy(IH) of neutrino masses. $m_{1,2,3}$ are real masses of $\nu_{1,2,3}$, $\phi_{1,2,3}$
the associated Majorana phases in diagonal form of $m_\nu$.  
For NH, $m_1=0$, $m_2=\sqrt{\Delta m^2_{21}}$, $m_3=\sqrt{|\Delta m^2_{32}|+\Delta m^2_{21}}$.
For IH,  $m_3=0$, $m_1=\sqrt{|\Delta m^2_{32}|-\Delta m^2_{21}}$, $m_2= \sqrt{|\Delta m^2_{32}|}$.
$x$ and $y$ are two real free parameters
to parametrize the mass matrix. (\ref{mixing1}) and (\ref{mixing2}) are valid for
large value of $y$,  i. e. for the case that cancellation in (\ref{0nubb-2}) is needed to  satisfy $0\nu\beta \beta$ constraint.

\begin{figure}[!htb]
\begin{tabular}{cc}
\includegraphics[scale=1,width=8cm]{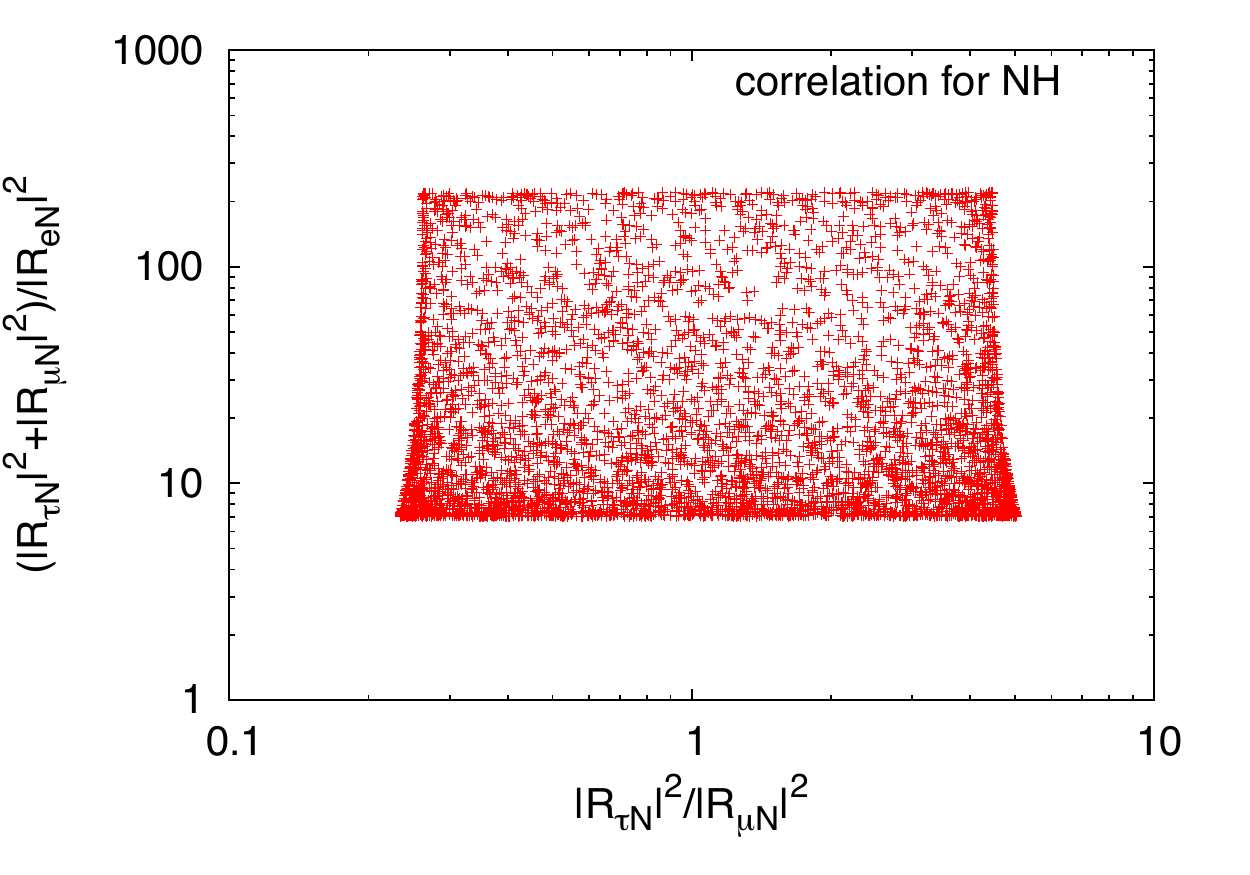}
\includegraphics[scale=1,width=8cm]{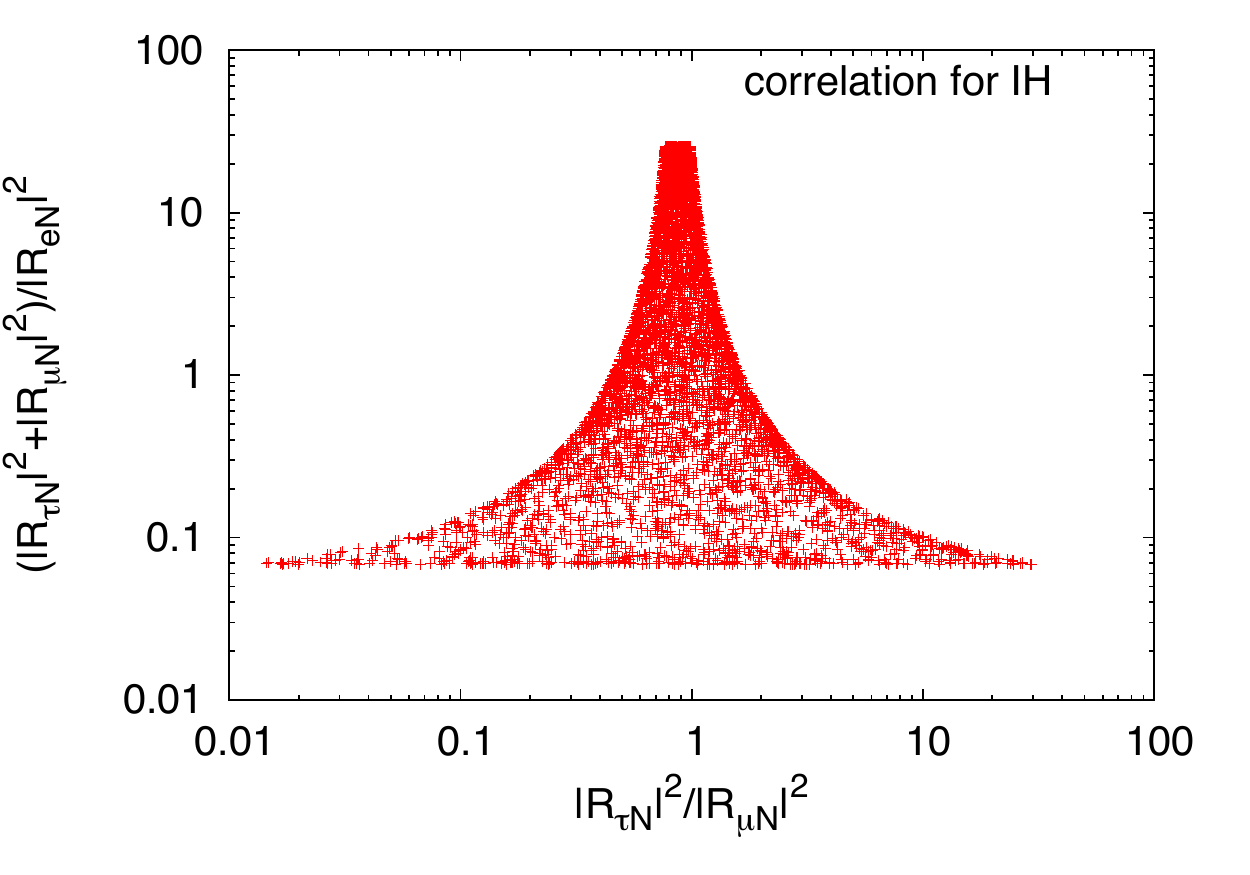}
\end{tabular}
\caption{\label{mixing-relation}
$(|R_{\tau N_1}|^2+|R_{\mu N_1}|^2)/|R_{e N_1}|^2$ versus $|R_{\tau N_1}|^2/|R_{\mu N_1}|^2$ 
for NH and IH respectively. 
}
\end{figure}

One can see in (\ref{mixing1}) and (\ref{mixing2}) that $|R_{l N_2}|^2 =|R_{l N_1}|^2$ is valid for all flavors of
neutrinos $\nu_{l=\tau,\mu,e}$ not just for $l=e$. This is one of the major properties of low energy see-saw model
if allowing large mixing of sterile neutrinos with active neutrinos.
Using (\ref{mixing1}) and (\ref{mixing2}) one can also show the correlation of $|R_{l N_1}|^2$
by varying the free Dirac phase in matrix $U$ and the Majorana phases $\phi_i$.
In Fig. \ref{mixing-relation} we plot the correlation of $(|R_{\tau N_1}|^2+|R_{\mu N_1}|^2)/|R_{e N_1}|^2$ 
versus $|R_{\tau N_1}|^2/|R_{\mu N_1}|^2$.
In our computation we use \cite{RPP}
\begin{eqnarray}
\sin^2 2\theta_{12}=0.846,  ~\sin^2 2\theta_{23}=0.999,~\sin^2 2\theta_{13} =0.093
\end{eqnarray}
and 
\begin{eqnarray}
\Delta m^2_{21}=7.53\times 10^{-5} ~\textrm{eV}^2,~
|\Delta m^2_{32}|=2.48\times 10^{-3} ~\textrm{eV}^2
\end{eqnarray}
For $\Delta m^2_{32}$ we have averaged two fit values for NH and IH~\cite{RPP}.
One can see in these plots that the mixings of sterile neutrino with $\nu_\tau$ and $\nu_\mu$ together are
always stronger than the mixing with $\nu_e$ for NH. For IH, $|R_{\tau N_1}|^2+|R_{\mu N_1}|^2$
can be larger than or smaller than $|R_{e N_1}|^2$ . On the other hand, the ratio between $|R_{\tau N_1}|^2$ and
$|R_{\mu N_1}|^2$ can be larger than or smaller than one for both NH and IH.

From the above discussions, one can see that a major implication of a low energy see-saw type model 
with two GeV scale sterile neutrinos and large mixings with active neutrinos is the relation of mixings, 
such as $|R_{l N_1}|^2=|R_{l N_2}|^2$ and the correlation shown in Fig. \ref{mixing-relation}.
For discussion of collider signatures in this low energy see-saw model, one should take these relations into account. 
However, as a first step towards this goal, we can discuss the signature of a single sterile neutrino with a mass at around 100 GeV.
The signature of low energy see-saw model  can be obtained by extending the discussion for a single sterile neutrino to two
sterile neutrinos and taking into account these relations among mixings described above. A further advantage of first discussing a single sterile neutrino
is that the case of a single sterile neutrino may also be valid if other particle or mechanisms, e.g. some scalar
particles and Type-II see-saw mechanism, are introduced. So a discussion on the collider signature of a single heavy sterile neutrino
is of interests for itself.  Needless to say, discussing the signature of
a heavy sterile neutrino together with signature of other particles, e.g. scalar particles in type-II see-saw mechanism,
is also of interests. In the present article, we are not going to elaborate on this topic. 
In the next section, we discuss the signature of a single sterile neutrino with a mass around 100 GeV at CEPC.
We come back to the signature of low energy see-saw model in later sections.
For previous works on
the signature of heavy sterile neutrino on $e^+ e^-$ collider, one can see a review in \cite{Antusch:2016ejd}. 
The present work give a discussion on the signature of heavy sterile neutrino on CEPC within the framework of low energy see-saw model
and differs from the previous works in these aspect. 

The mixings of sterile neutrinos with active neutrinos are also
subject to the indirect constraints from tests of lepton universality,
lepton flavor violation processes and electroweak precision
measurements~\cite{Antusch:2006vwa,Antusch:2008tz,Ibarra:2011xn,
Drewes:2015iva,deGouvea:2015euy,Fernandez-Martinez:2016lgt}.
For the heavy neutrino masses of order of $100$GeV,
$|R_{eN}|^2$, $|R_{\mu N}|^2$ and $|R_{\tau N}|^2$ are constrained
to be ${\cal O} (10^{-3})$ mainly by the lepton flavor conserved decays
of charged leptons, mesons, $W$, and $Z$.
The combination $|R_{\mu N}^\ast R_{eN}|$ are stringently constrained
to be order of $10^{-5}$ from the upper bounds of $\mu \to e\gamma$
and $\mu - e$ conversion.
These indirect constaints are complementary to the probing of
heavy sterile neutrino at $e^+ e^-$ colliders.
As will be shown in later sections,
CEPC will probe $|R_{lN}|^2$ to $10^{-5}-10^{-6}$ and will have a better
sensitivity than these indirect constraints at present.

\bigskip
\section{Production and decay of a heavy sterile neutrino}
In this section, we discuss the production of a single heavy sterile neutrino at CEPC and
its decay.  CEPC under proposal plans to run electron positron collision at a
center-of-mass energy around 240 GeV and aims at obtaining an integrated luminosity up to $5$ ab$^{-1}$
with two interaction points and ten years of operation.

\begin{figure}[!htb]
\includegraphics[scale=1,width=12cm]{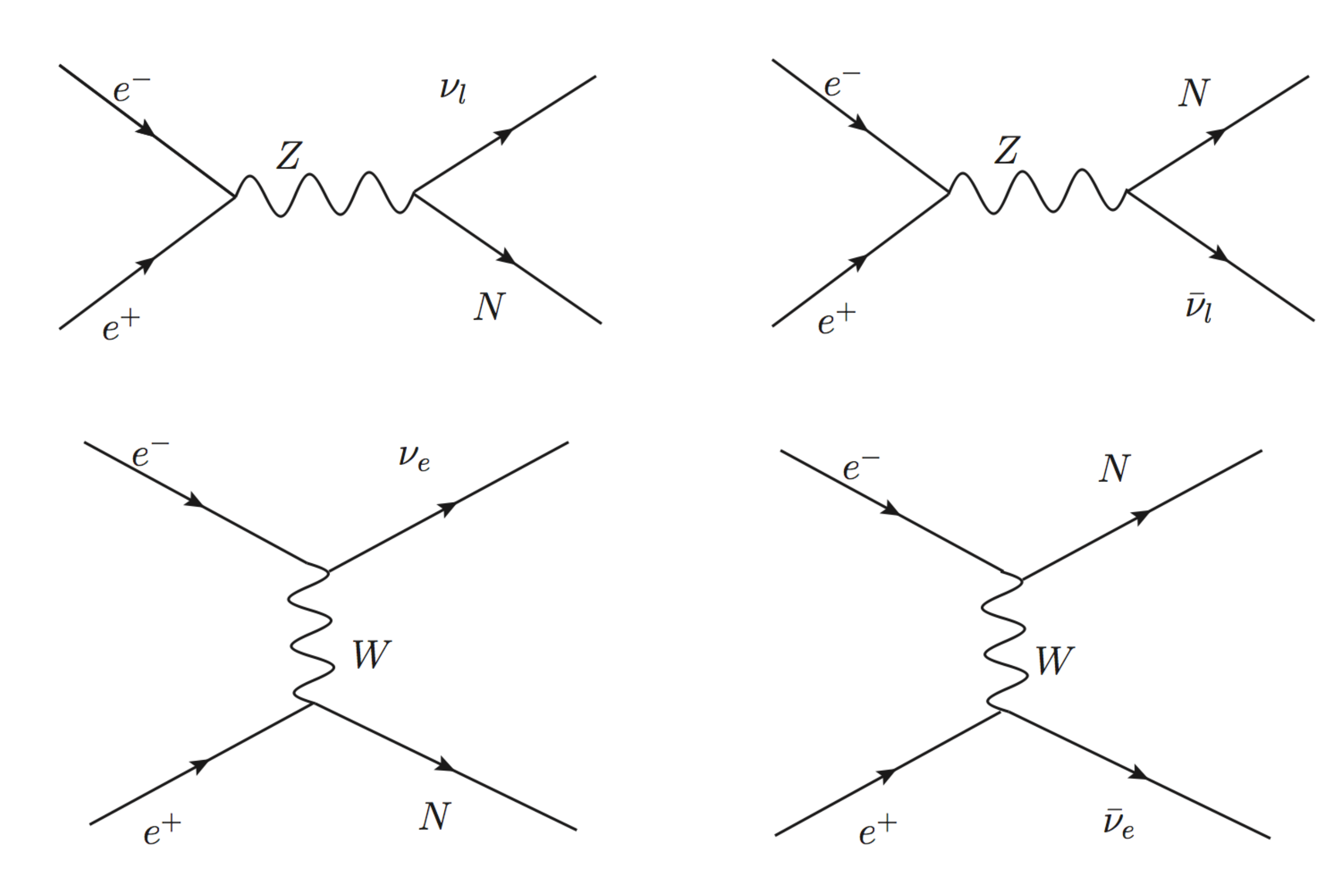}
\caption{\label{feynmandiagram}
Feynman diagrams
}
\end{figure}

The Feynman diagrams of the production of a heavy Majorana-type sterile neutrino, $N$, 
are shown in Fig.~\ref{feynmandiagram}.  
For simplicity, the heavy neutrino index $j$ will be suppressed in discussion for a single heavy sterile neutrino.
The leading contribution to $N$ production is the process $e^+ e^- \to N \nu_l({\bar \nu}_l)$,
the SM process $e^+ e^- \to \nu_l {\bar \nu}_l$ with $\nu_l$ or ${\bar \nu}_l$ replaced by $N$
 via its mixing with $\nu_l$. Because of the Majorana nature of N, 
it can mix with both of $\nu_l$ and ${\bar \nu}_l$ with the same strength of mixing and can be produced via both of these mixings.
These two possibilities are shown in the left and right panels in Fig. \ref{feynmandiagram}.
As one can see in upper panels of Fig. \ref{feynmandiagram},  the production of $N$ can be mediated by 
a $Z$ boson in s-channel with all type of neutrinos $\nu_l({\bar \nu}_l)$ in final state. 
$N$ production can also be mediated by a $W$ boson in t-channel with $\nu_e({\bar \nu}_e)$
in final state, as can be seen in lower panels in Fig. \ref{feynmandiagram}. 
For the same strength of mixings,  the t-channel process has a cross section two order of magnitude larger 
than the s-channel process and hence
has a better sensitivities for the mixing $R_{eN}$.

\begin{figure}[!htb]
\centerline{\includegraphics[width=8cm] {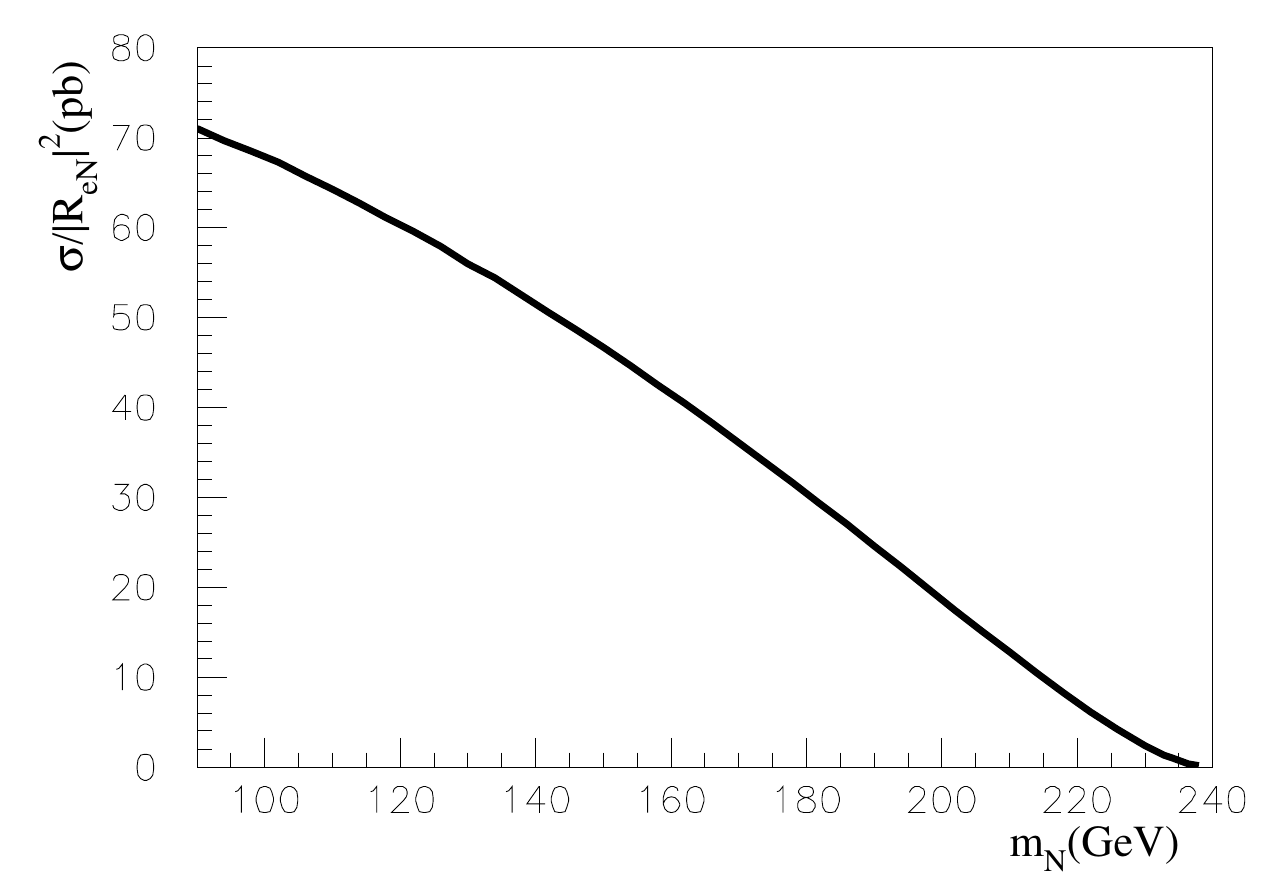}
\includegraphics[width=8cm] {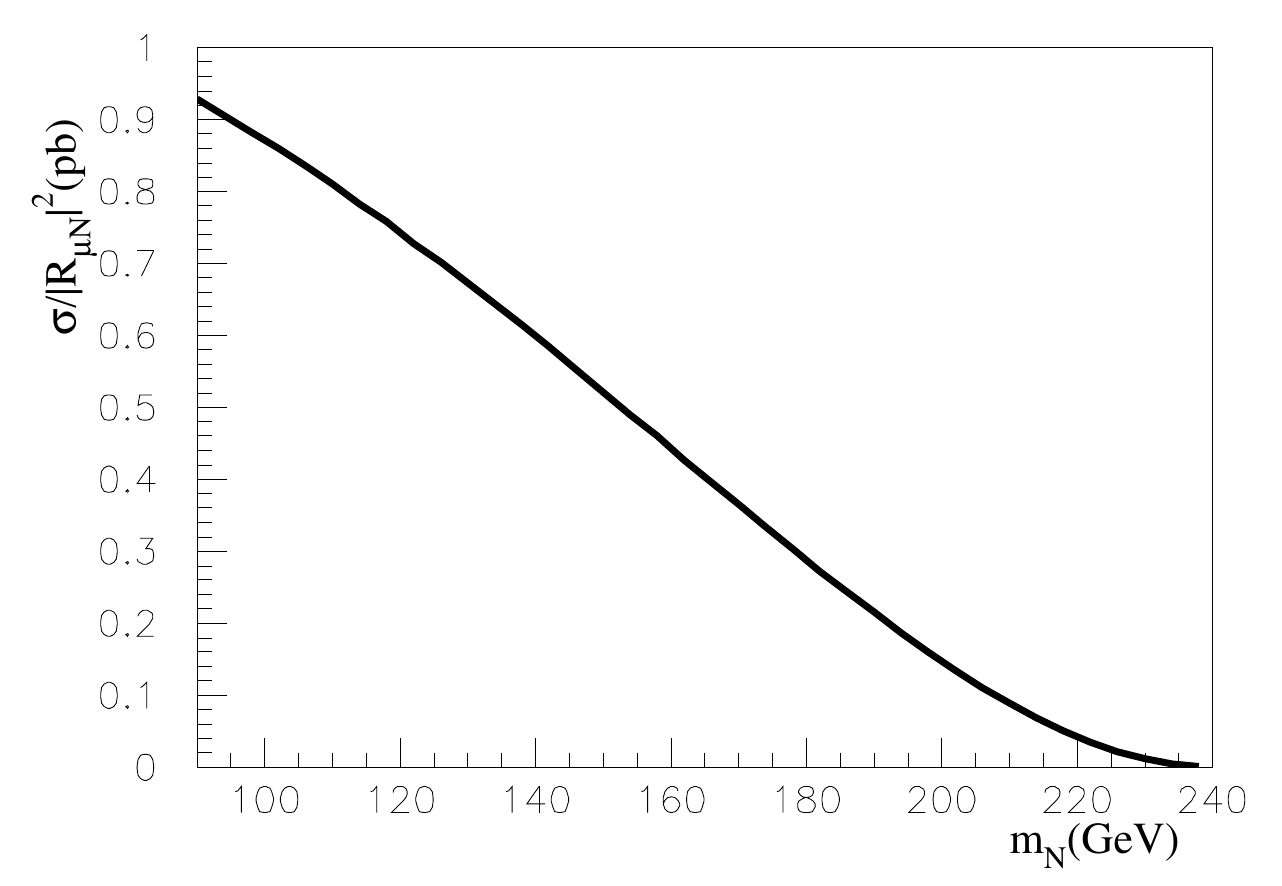}}
\caption{\label{xsection}
$e^+ e^- \to N \nu$ cross section at $\sqrt{s}=240 {\rm GeV}$
with only a single $R_{eN}$ mixing (left) and $R_{\mu N}$ mixing (right).
}
\end{figure}

We calculate the tree-level $e^+ e^- \to N \nu_l$ cross sections
with MadGraph~\cite{Alwall:2014hca}
and implement the heavy neutrino interactions
in FeynRules~\cite{Alloul:2013bka}
with the Universal FeynRules Output (UFO)~\cite{Degrande:2011ua}
format for the model.
The results are shown in Fig.~\ref{xsection}. 
For a heavy neutrino of about $100 {\rm GeV}$,
the production cross section of $\sigma/|R_{eN}|^2$ and $\sigma/|R_{\mu N}|^2$
can reach $\sim 60 {\rm pb}$ and $\sim 0.8 {\rm pb}$
for only a single $R_{eN}$ mixing or $R_{\mu N}$ mixing, respectively.

Mixing of sterile neutrino $N$ with active neutrinos can lead to decay of $N$.
For $m_N$,  the mass of $N$,  much smaller than $m_W$, the mass of W boson, the leading decays of $N$
are tree-level three-body decays mediated by off-shell W or Z bosons. Some three-body decay channels of $N$ are quite simple.
For example, $N \to e^- \mu^+ \nu_\mu$ is mediated by an off-shell W boson and is similar to  $\mu \to \nu_\mu e {\bar \nu}_e$,
the leptonic decay of $\mu$, except with the presence of a mixing factor $|R_{eN}|^2$ in decay rate.
Some decay channels, e.g. $N \to \nu_e e^- e^+$, can be mediated by both off-shell W and Z bosons.
But it does not introduce complications in the decay rate. The results are presented in (\ref{decayR1-1}),
(\ref{decayR1-2}), (\ref{decayR1-3}), (\ref{decayR1-4}), (\ref{decayR1-5}), (\ref{decayR1-6}), (\ref{decayR1-7}) in Appendix.

For $m_N$ much greater than $m_W$ and $m_Z$, the leading decay of $N$ are two-body decays, $N \to l^\pm W^\mp$ and $N\to \nu({\bar \nu}) Z$.
For $m_N$ greater than $m_H$, the mass of Higgs boson, $N$ can also decay to $H$ via $N \to \nu({\bar \nu})H$.
The partial decay widths of the heavy neutrino can be written as~\cite{Atre:2009rg,
delAguila:2008cj, Liao1, Banerjee:2015gca}
\begin{eqnarray}
\Gamma(N \to l^- W^+) &=& \frac{g^2}{64\pi} |R_{lN}|^2 \frac{m_N^3}{m_W^2}
  (1-\mu_W)^2 (1 + 2\mu_W)  \label{decay01}\\
\Gamma(N \to \nu Z) &=& \frac{g^2}{64\pi} |R_{lN}|^2 \frac{m_N^3}{m_W^2}
  (1-\mu_Z)^2 (1 + 2\mu_Z)  \label{decay02}\\
\Gamma(N \to \nu H) &=& \frac{g^2}{64\pi} |R_{lN}|^2 \frac{m_N^3}{m_W^2}
  (1-\mu_H)^2 \label{decay03}
\end{eqnarray}
with $\mu_i=m_i^2/m_N^2$ $(i = W, Z, H)$.  $W$, $Z$ or $H$ eventually decay to fermions.
Hence the decay rate to a specific three-body final states can be calculated using (\ref{decay01}-\ref{decay03})
and the branching ratio of $W$, $Z$ or $H$ to a specific fermion pair. For example,
$\Gamma(N \to e^- \mu^+ \nu_\mu)$ is obtained using $\Gamma(N\to e^- W^+)$ and
$Br(W\to \mu^+ \nu_\mu)$ as $\Gamma(N \to e^- \mu^+ \nu_\mu)=\Gamma(N\to e^- W^+) Br(W^+\to \mu^+ \nu_\mu)$
where $Br(W^+\to \mu^+ \nu_\mu)$ is the branching ratio of $W^+\to \mu^+ \nu_\mu$ decay.

\begin{figure}[!htb]
\begin{center}
\includegraphics[width=10cm]{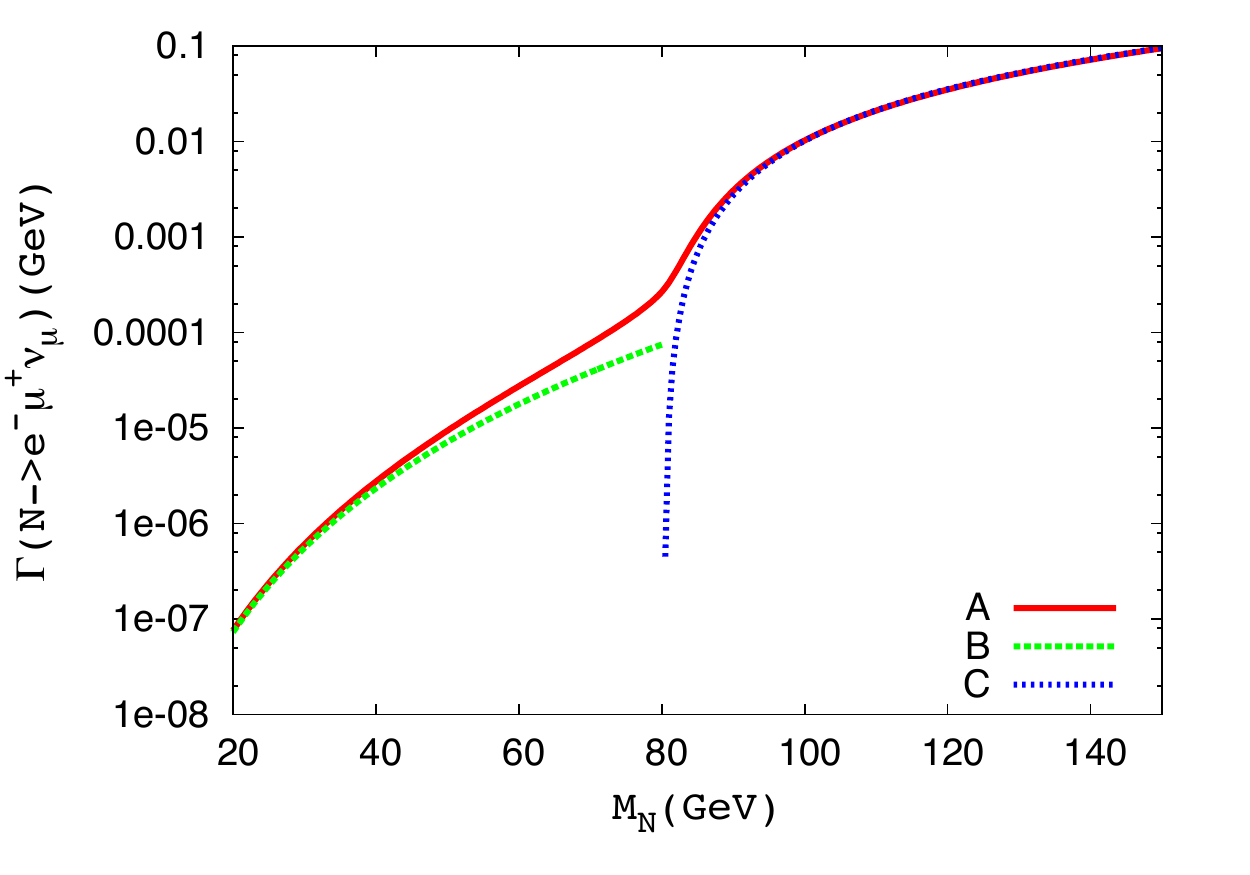}
\end{center}
\caption{\label{decayrate01}
Decay rate of $N \to e^- \mu^+ \nu_\mu$ versus $m_N$ with $|R_{e N}|^2=1$.
Line A: calculated using (\ref{decayR1}); Line B: calculated using $\Gamma=G_F^2 m_N^5/(192 \pi^3)$ in (\ref{decayR1-1})
up to $m_N< m_W$; Line C: calculated using $\Gamma=\Gamma(N\to e^- W^+) Br(W^+\to \mu^+ \nu_\mu)$
with $m_N > m_W$.}
\end{figure}

For more general value of $m_N$,  in particular for $m_N\approx m_{W,Z,H}$, the above formulas
are not good approximations. Decay rate in more general cases can be calculated by carefully including the
propagators of $W$, $Z$ and $H$ bosons into calculation.  Four-momentum of the mediated boson can
be on-shell for general cases.  We take this fact into account and calculate the tree-level decay rate of $N$ decays with
three fermions in final state. In Appendix, we present in detail our result of calculation. One can see that
for most cases the decay rate can be obtained as an analytic function of $m_N$ and the masses and widths of bosons.
The most complicated case appears for $N \to l^- l^+ \nu_l$ and $N \to l^- l^+ {\bar \nu}_l$ channels
for which $W$ and $Z$ bosons can all mediate. For this particular process, a function $F_S$, shown in (\ref{decayR3}) and
(\ref{FS}), appears which cannot be obtained as an explicit analytic function of $m_N$ and the boson masses. 
In our analysis we compute $F_S$ numerically.  

As an example, we compare in Fig. \ref{decayrate01} the result computed using analytic formula (\ref{decayR1})
with known results in low energy region $m_N \ll m_W$ and in high energy limit $m_N > m_W$. 
$\Gamma(N\to e^- W^+) $ is calculated using (\ref{decay01}).
$Br(W^+\to \mu^+ \nu_\mu)$  is taken as $Br=0.108$\cite{RPP}. We can see that in the low energy limit the
decay rate agrees with the expected result of tree-level three-body decay. In the high energy limit it agrees with the expectation that it is dominated
by the on-shell $N \to e^- W^+$ decay with a subsequent $W ^+\to \mu^+ \nu_\mu$ decay. In region around $m_W$,
(\ref{decayR1}) gives a smooth transition from low energy behavior to high energy behavior.  As a comparison, the
result calculated using the two body decay $N\to e^- W^+$ drops down to zero as $m_N$ approaches $m_W$ from
above and is certainly not correct at around threshold. The result given by (\ref{decayR1}) takes into account the
contribution of off-shell boson and removes the ill-behavior at around $m_N \sim m_W$.
The plot demonstrates that results presented in
Appendix are better to use for studying the signals of sterile neutrino. 
Tree-level three-body decay rates for general mass $m_N$, presented in appendix, are some of the new results
of the present article.

\bigskip
\section{Signal of a heavy sterile neutrino and background}

In this section, we study the process 
\begin{eqnarray}
e^+ e^- \to N \nu, N{\bar \nu} \to l j j \slashed{E}, \label{process}
\end{eqnarray}
the signal of sterile neutrino $N$ due to this process and the associated background.

We simulate the signal and background events
with MadGraph~\cite{Alwall:2014hca},
and have done the showing and hadronization
by using Pythia6~\cite{Sjostrand:2006za}.
The results are passed through PGS4~\cite{PGS4}
for fast detector simulation. 

At CEPC with $\sqrt{s}=240$ GeV,
we adopt the basic cuts (BC) for lepton and jets to select the events,
\begin{eqnarray}
p_T^l &>& 10 {\rm GeV}, |\eta^l|<2.5, \Delta R_{ll} > 0.4, \\
p_T^j &>& 10 {\rm GeV}, |\eta^j|<2.5, \Delta R_{jj} > 0.4,
  \Delta R_{lj} > 0.4.
\end{eqnarray}

The main backgrounds for the process (\ref{process})
are $W$ pair production, $e^+ e^- \to W^+ W^-$, 
with one $W$ decaying leptonically and the other $W$ decaying hadronically,
and single $W$ production, which decays leptonically.
In order to suppress the backgrounds,
we set the selection cuts (SC)~\cite{Atre:2009rg,Banerjee:2015gca},
\begin{eqnarray}
|M(l,\slashed{E}) - m_W| > 20 ~{\rm GeV},
\label{cut1}
 \end{eqnarray}
  and
 \begin{eqnarray}
|M(l,j_1,j_2) - m_N| < 20 \hspace{2mm} {\rm or}\hspace{2mm} 10 ~{\rm GeV}.
\label{cut2}
\end{eqnarray}
Cut (\ref{cut1}) is used to exclude background events
coming from the decay of on-shell W boson
in the background processes.  Cut (\ref{cut2}) selects events coming from
the decay of on-shell $N$ and is used to
increase the significance of signal to background ratio. 

In Table \ref{cuts_efficiency} we show the efficiency of the cuts
for both $l=e$ and $l=\mu$ channels. 
After adding the SC, the signals are survived,
but the backgrounds drop several order of magnitude.

\begin{table}[!htb]
\caption{\label{cuts_efficiency}
The cross sections (unit fb) of signal (upper line) after imposing various cuts (a, b, c, d, e) sequentially,
the background (lower line)
and the significance after cuts
with integrated luminosity of $500{\rm fb}^{-1}$.
Cuts (a) $p^{j,l}_T > 1$GeV, (b) $p^{j,l}_T > 10$GeV,
(c) $|M(l,\slashed{E}) - m_W| > 20$GeV, 
(d) $|M(l,j_1,j_2) - m_N| < 20$GeV, (e) $|M(l,j_1,j_2) - m_N| < 10$GeV.
}
\begin{tabular}{|c|c|c|c|c|c|c|c|}
\hline
& parameters & +cuts (a) & +cuts (b) & +cuts (c) & +cuts (d) &
 +cuts (e) & significance \\
\hline
A & $m_N=150$GeV, & $2.14$ & $2.04$ & $1.56$ & $1.56$ & $1.55$ & $11.2$ \\
& $R_{\mu N}=0.1$ & $2.31\times 10^3$ & $2.20\times 10^3$ & $52.4$ & $16.3$ &
  $8.05$ & \\
\hline
B & $m_N=150$GeV, & $7.63$ & $7.30$ & $5.61$ &  $5.60$ &  $5.60$ &
  $18.8$  \\
& $R_{eN}=0.02$ & $2.52\times 10^3$ & $2.37\times 10^3$ & $0.195\times 10^3$ &
 $76.6$ & $38.8$ & \\
\hline
C & $m_N=90$GeV, & $10.8$ & $4.98$ & $1.56$ &  $1.55$ &  $1.55$ & $13.4$  \\
& $R_{eN}=0.015$ & $2.52\times 10^3$ & $2.37\times 10^3$ & $0.195\times 10^3$ & 
 $16.8$ & $5.14$ & \\
\hline
D & $m_N=214$GeV, & $0.852$ & $0.827$ & $0.243$ &  $0.242$ &  $0.241$ &
 $1.75$  \\
& $R_{eN}=0.015$ & $2.52\times 10^3$ & $2.37\times 10^3$ & $0.195\times 10^3$ & 
 $24.9$ & $9.26$ & \\
\hline
E & $m_N=230$GeV, & $0.194$ & $0.188$ & $0.160$ &  $0.160$ &  $0.160$ &
 $2.76$  \\
& $R_{eN}=0.015$ & $2.54\times 10^3$ & $2.39\times 10^3$ & $0.197\times 10^3$ &
 $4.14$ & $1.49$ & \\
\hline
\end{tabular}
\end{table}

\begin{figure}[!htb]
\includegraphics[width=8cm]{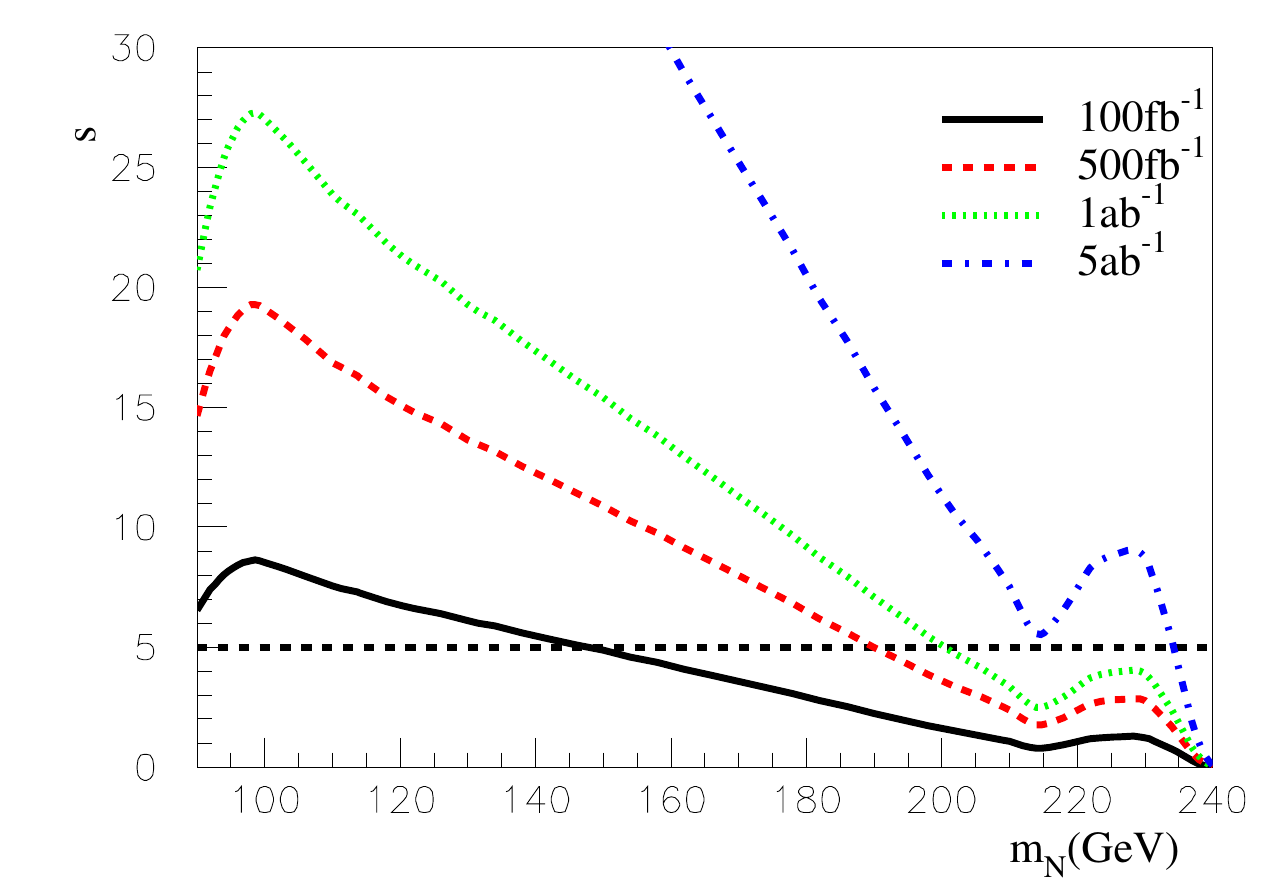}
\includegraphics[width=8cm]{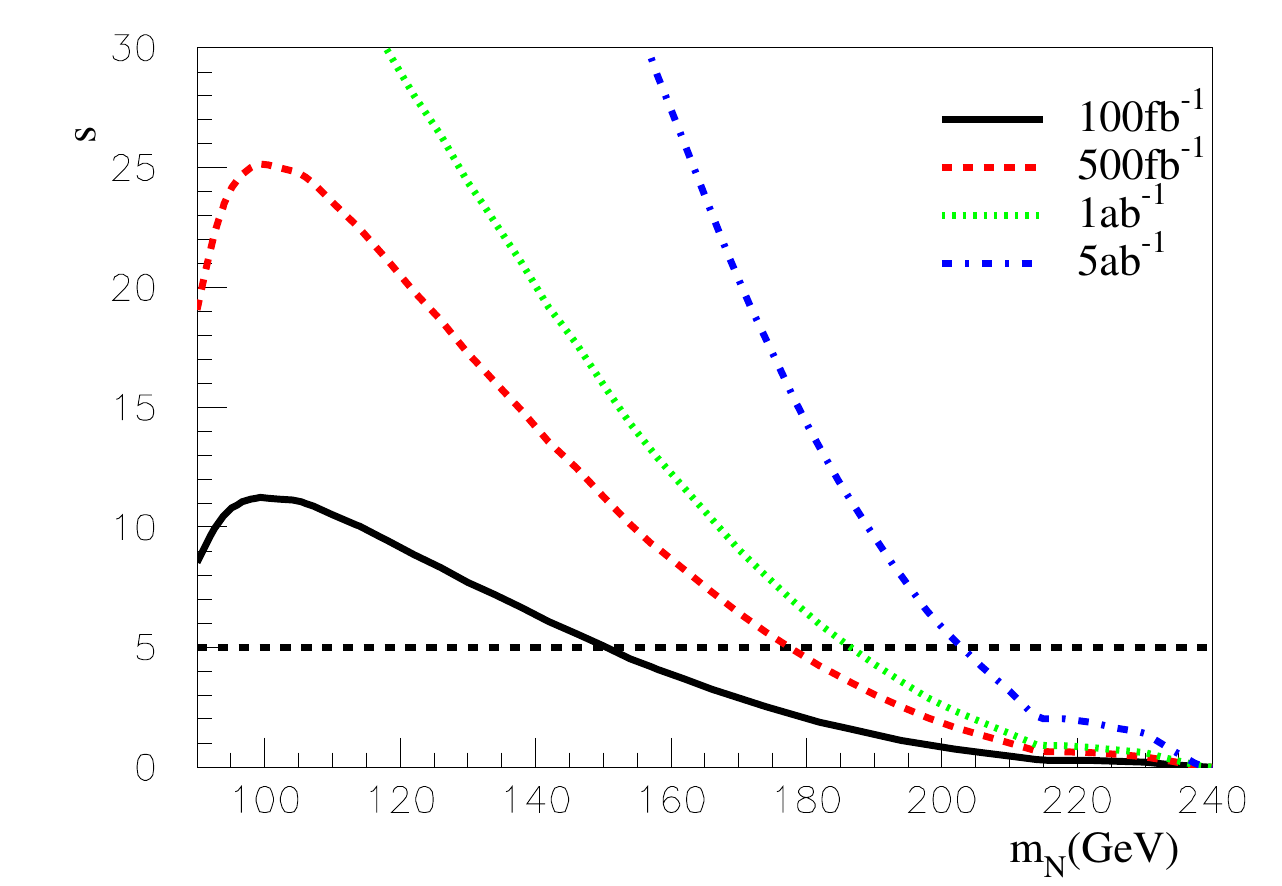}
\caption{\label{significance}
The significance for $l=e$ (left) with $R_{eN}=0.015$
and $l=\mu$ (right) with $R_{\mu N}=0.1$.
The curves in each plot from up to down correspond to
the integrated luminosities $5{\rm ab}^{-1}$, $1{\rm ab}^{-1}$,
$500{\rm fb}^{-1}$ and $100{\rm fb}^{-1}$.
}
\end{figure}

We define the significance $s$ as
\begin{eqnarray}
s = \frac{{\cal N}_s}{\sqrt{{\cal N}_s + {\cal N}_b}},
\end{eqnarray}
where ${\cal N}_s$ and ${\cal N}_b$ are the event number
of signal and background respectively.
In Fig. \ref{significance} we plot the significance $s$ versus $m_N$
for $l=e$ with $R_{eN}=0.015$ and $l=\mu$ with $R_{\mu N}=0.1$, respectively.
For the integrated luminosity of 100 ${\rm fb}^{-1}$,
a heavy neutral neutrino with mass in the range of
$90 {\rm GeV} \le m_N \le 146 {\rm GeV}$
for the mixing $R_{eN}=0.015$ is promised to be discoverd in $l=e$ channel,
and 
$90 {\rm GeV} \le m_N \le 150 {\rm GeV}$
for the mixing $R_{\mu N}=0.1$ is promised to be discoverd in $l=\mu$ channel.
For the integrated luminosity of $5{\rm ab}^{-1}$,
the maximal values of heavy neutrino mass can be
$235 {\rm GeV}$ and $205 {\rm GeV}$
for $l=e$ and $l=\mu$ channel, respectively.
One can see that there is a quick drop for heavy neutrino
with mass $\lsim 100{\rm GeV}$
for both of $l=e$ and $l=\mu$. This is because 
for the decay of N of a mass  $\lsim 100$ GeV, the lepton
in $N \to l W \to l j j$, a decay chain with an almost on-shell W,
does not have enough energy and the $p_T$ of $l$ can not be large.
This effect of cut on $p_t$ of $l$ can be seen
in Table.~\ref{cuts_efficiency} C. 
In Fig. \ref{significance}, one can also see that there
is a small peak for a heavy neutrino with mass around $230{\rm GeV}$.
This is because of the cut $|M(l,\slashed{E}) - m_W| > 20{\rm GeV}$
to the signal as shown in Table.~\ref{cuts_efficiency} E.
Compared with the case of $m_N=214$GeV in Table.~\ref{cuts_efficiency} D,
a heavier neutrino with mass of $230$GeV tends to move more slowly
in the center of mass system of colliding $e^+e^-$,
and it decays to a charged lepton
which distributes more uniformly in all directions.
More importantly, the light neutrino, produced together
with the heavier sterile neutrino with a mass of $230$GeV,
becomes quite soft (with an energy $\approx 9.8$ GeV).
Then, the invariant mass of the light neutrino
and charged lepton $M(l,\slashed{E})$ will distribute more evenly.
Consequently, the cut $|M(l,\slashed{E}) - m_W| > 20{\rm GeV}$
does not hurt the signal as much as in the case of $m_N=214$GeV,
as shown in Table.~\ref{cuts_efficiency} D and E.
This can be verified if the cut $(c)$ is
changed to $|M(l,\slashed{E}) - m_W| > 10\, (30){\rm GeV}$,
the signal cross section is changed to $0.188\, (0.103)$fb, respectively.
On the other hand, the background cross section reduces significantly
with $m_N=230$GeV in Table.~\ref{cuts_efficiency} E after adding all the cuts.

\begin{figure}[!htb]
\includegraphics[width=8cm]{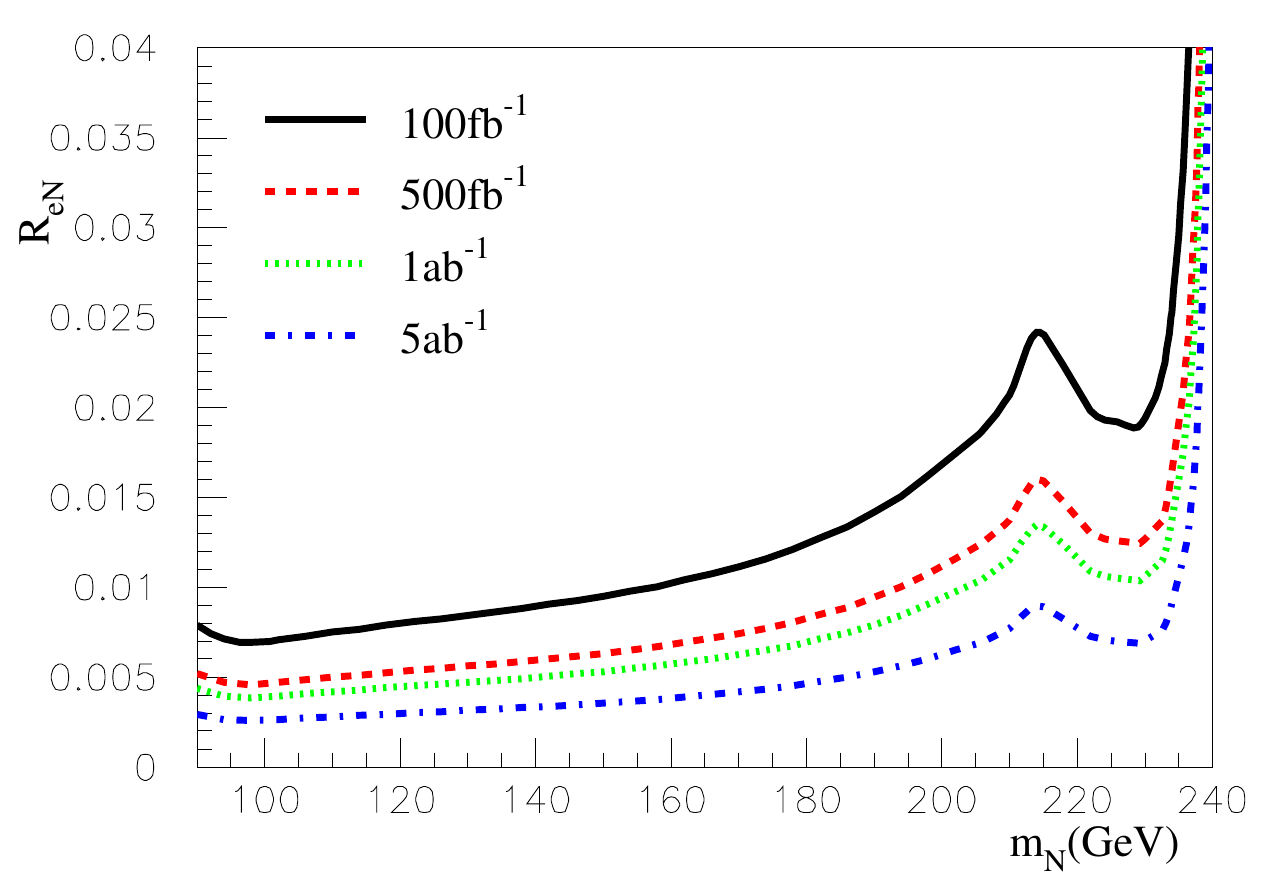}
\includegraphics[width=8cm]{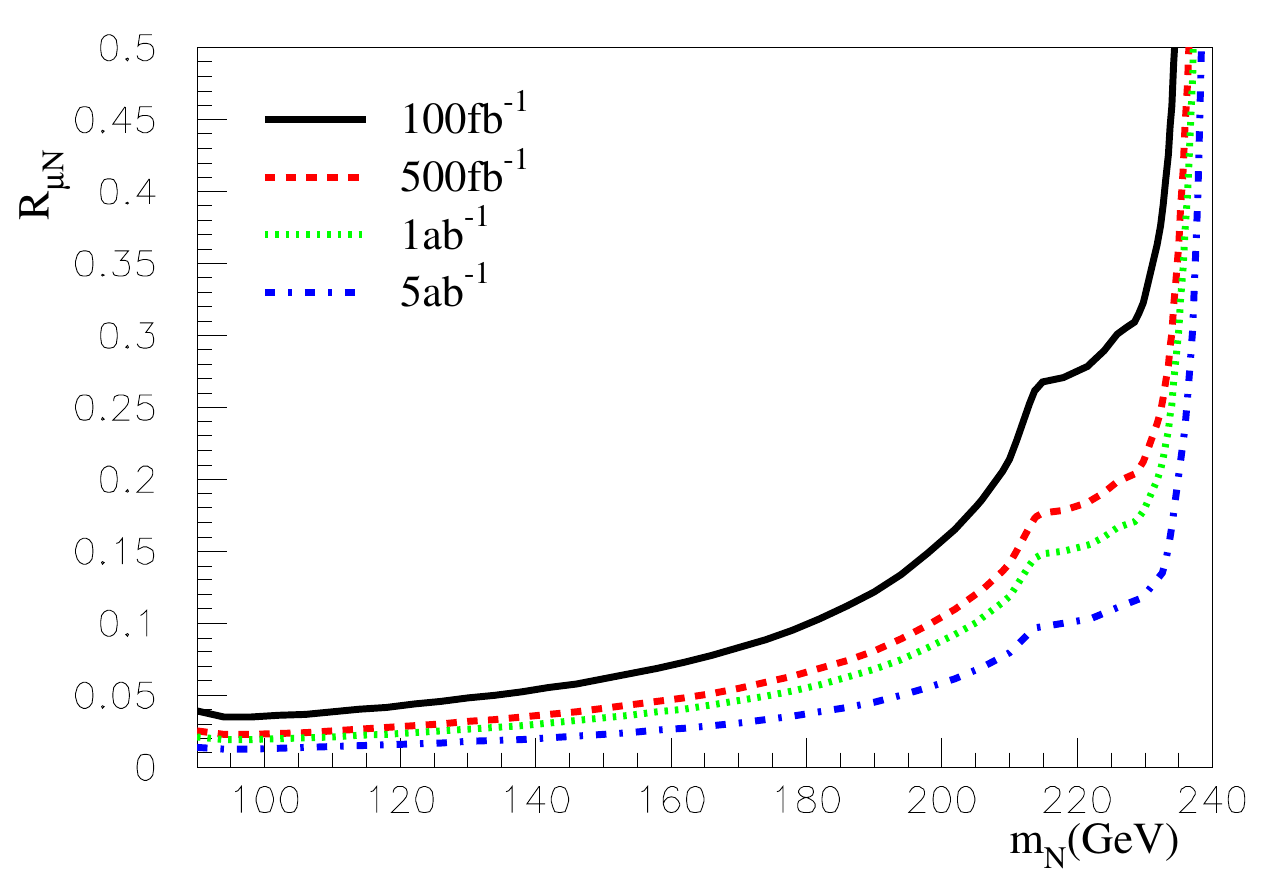} \\
\includegraphics[width=8cm]{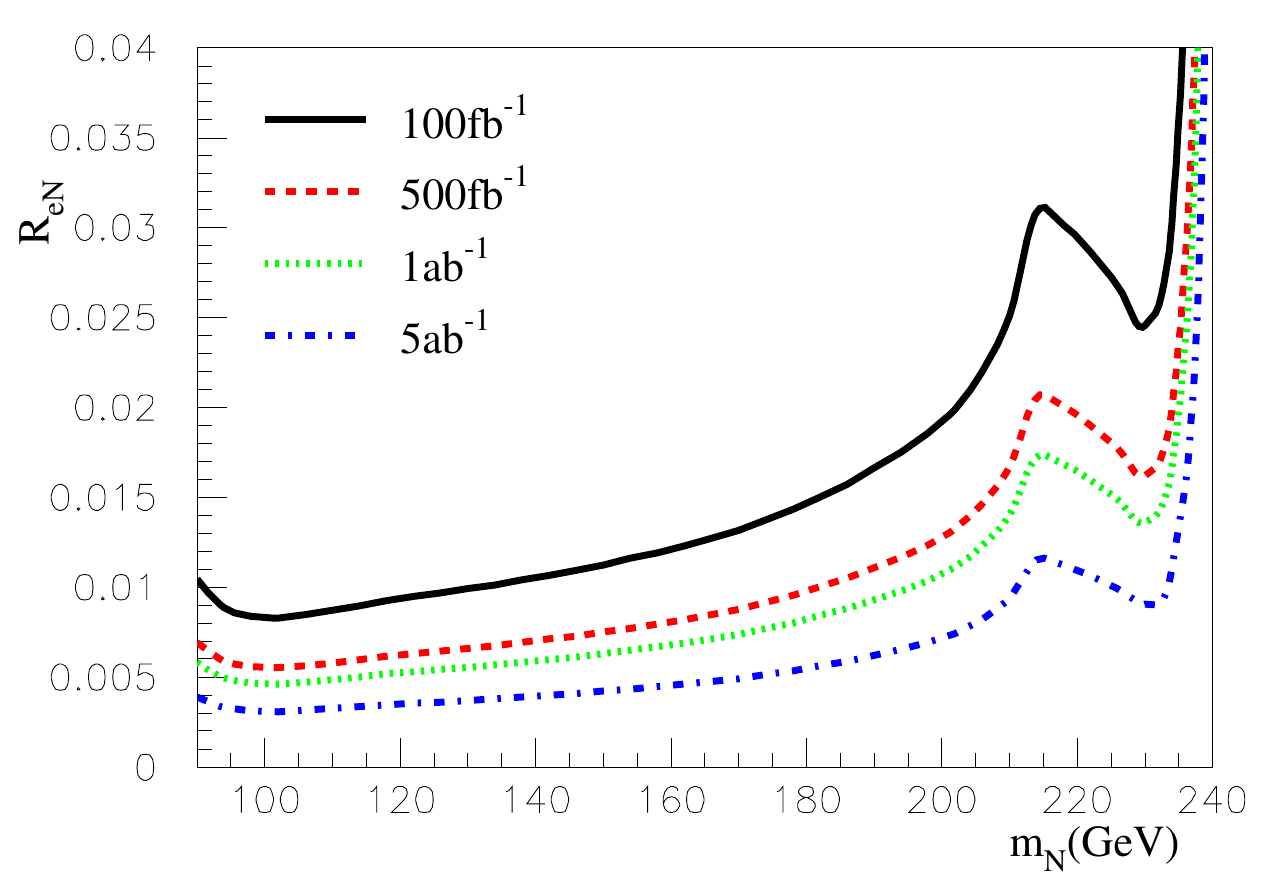}
\includegraphics[width=8cm]{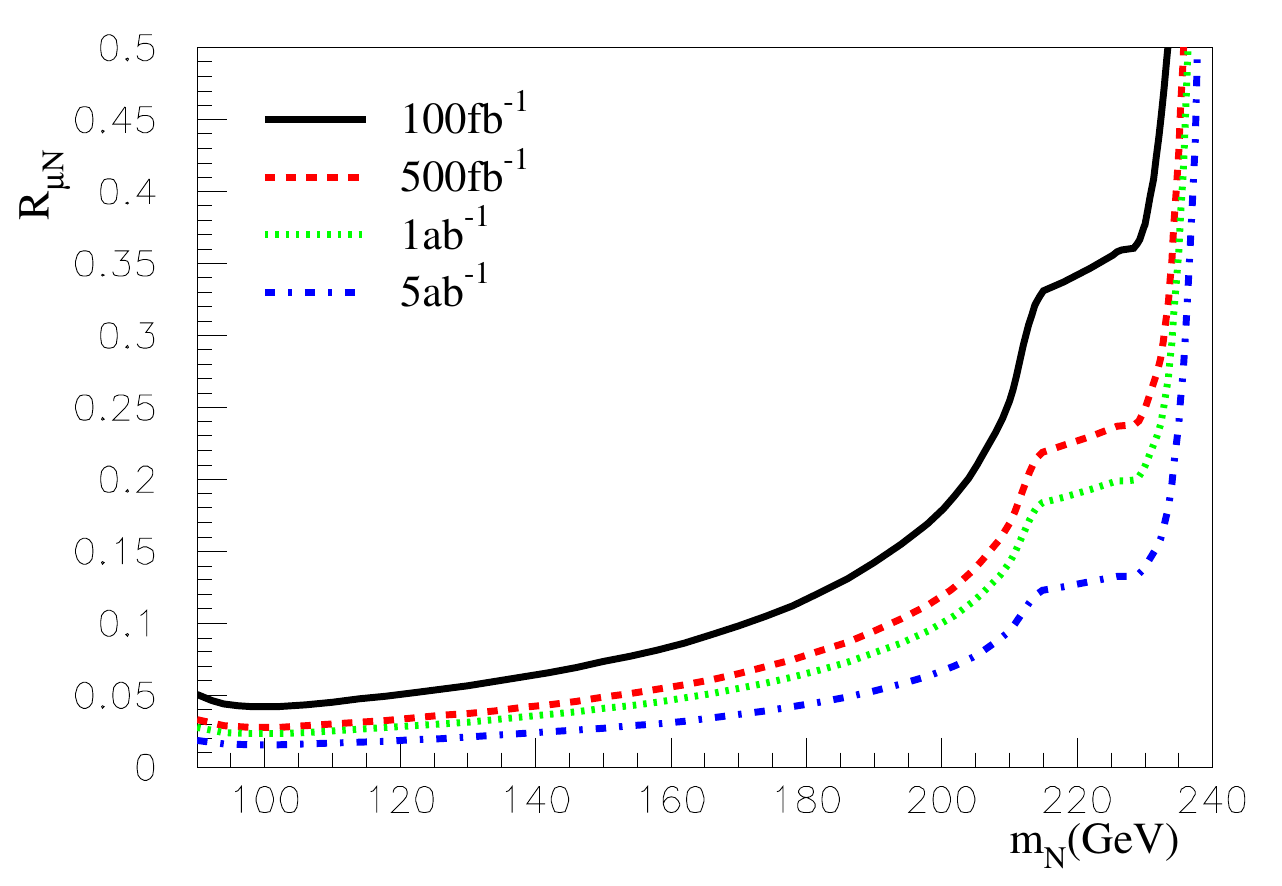}
\caption{\label{RlN}
Sensitivity to $R_{lN}$ ($l=e, \mu$) with significance $s=5$.
The upper plots are for SC $|M(l,j_1,j_2) - m_N| < 10{\rm GeV}$
and the lower plots are for SC $|M(l,j_1,j_2) - m_N| < 20{\rm GeV}$.
The curves in each plot from up to down correspond to
integrated luminosities $100{\rm fb}^{-1}$, $500{\rm fb}^{-1}$, $1{\rm ab}^{-1}$ and $5{\rm ab}^{-1}$.
}
\end{figure}

In Fig. \ref{RlN} we plot the potential of probing $R_{lN}$
for a fixed significance $s=5$
with the integrated luminosities $5{\rm ab}^{-1}$, $1{\rm ab}^{-1}$,
$500{\rm fb}^{-1}$ and $100{\rm fb}^{-1}$ at CEPC.
Using SC $|M(l,j_1,j_2) - m_N| < 10{\rm GeV}$,
in $l=e$ channel, a heavy neutrino mass of $120 {\rm GeV}$
with $R_{eN}=0.0080$ can be discovered
for the integrated luminosities $100{\rm fb}^{-1}$,
and for $5{\rm ab}^{-1}$,
the mixing as low as $R_{eN}=0.0030$ for the same mass can be probed.
In $l=\mu$ channel, the heavy neutrino of the same mass
with $R_{\mu N}=0.043$ can be discovered for $100{\rm fb}^{-1}$,
and $R_{\mu N}=0.016$ for $5{\rm ab}^{-1}$.
We can have similar results for SC $|M(l,j_1,j_2) - m_N| < 20{\rm GeV}$,
but the corresponding mixings are a little bigger.

\bigskip
\section{Signal of low energy see-saw model}
In this section we discuss the signature of low energy see-saw model
with two heavy sterile neutrinos of mass around 100 GeV. 

As discussed in previous section, in the case of large mixing
of heavy sterile neutrinos
with active neutrinos, not only the masses of these two sterile neutrinos
are (quasi)degenerate but also 
the mixing has a simple relation $R_{lN_2}=\pm i R_{lN_1}$
as shown in (\ref{mixing1}) and (\ref{mixing2}).
So the signature of the low energy see-saw model discussed here
is just the double of the result presented
for a single heavy sterile neutrino, except that we need to take into account
the correlation of the mixing $R_{lN}$ for different $l$
in the low energy see-saw model, as shown in Fig. \ref{mixing-relation}.

We calculate the signal of $e^+ e^- \to \nu ljj$ events and the related background for $l=e,\mu,\tau$ separately.
Then we calculate the significance of $e^+ e^- \to \nu ljj$ events for $l=e,\mu,\tau$ separately.
The total significance is defined as the square root of
the sum of the squares of the significances of signals
of $l=e$, $l=\mu$ and $l=\tau$,
which we call $e+\mu+\tau$ significance.
Similarly, we can define $e+\mu$ significance
which include signals of  $l=e$ and $l=\mu$.
For simplicity, we assume $100\%$ efficiency of the identification of $\tau$ lepton.
A realistic efficiency can be put into analysis without difficulty and would give rise to
a result in-between the lines of $e+\mu+\tau$ significance and $e+\mu$ significance
presented in figures below. 

\begin{figure}[!htb]
\centerline{ \includegraphics[width=8cm]{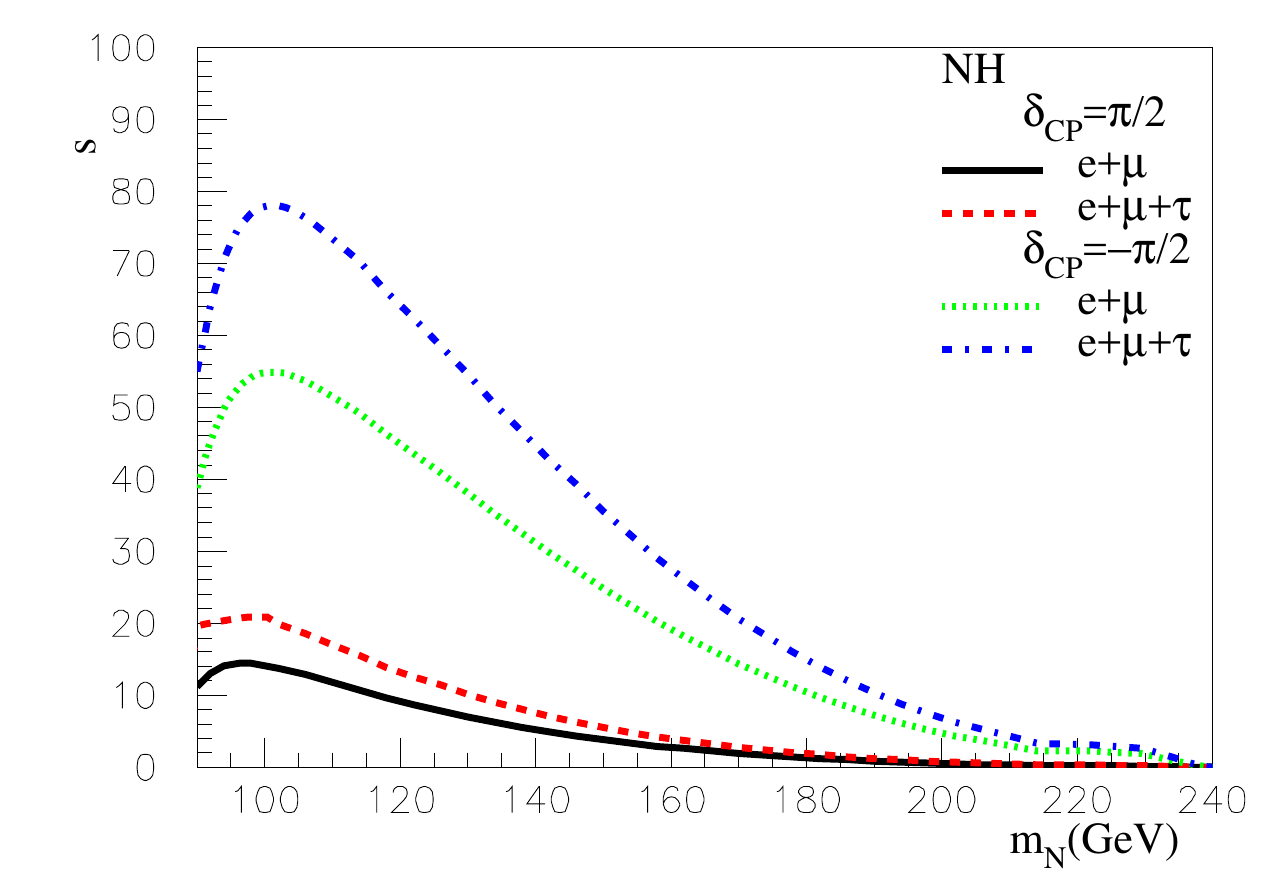}
 \includegraphics[width=8cm]{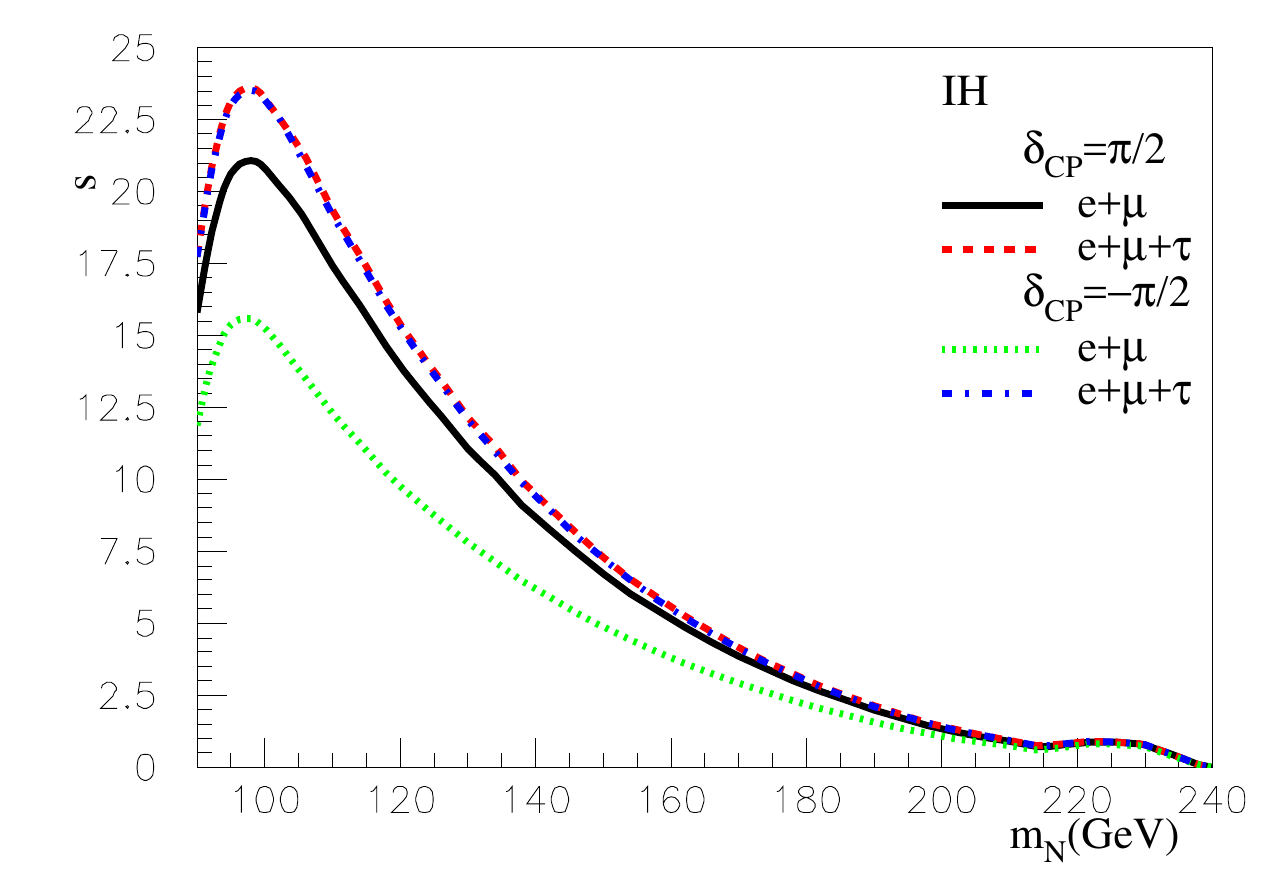}}
\caption{\label{s}
The significance $s$ vs $m_N$ for NH (left) and IH (right)
with integrated luminosity $500{\rm fb}^{-1}$.
We choose $e^y=5000$, $\delta_{\rm CP}=\pm \pi/2$,
$\phi_1=\phi_2=\phi_3=0$ for NH
(the largest eigenvalue of the matrix of neutrino Yukawa couplings
defined in Ref.~\cite{Ibarra:2011xn} is $0.00495\times \sqrt{m_N}$),
and $e^y=1000$, $\delta_{\rm CP}=\pm \pi/2$,
$\phi_1=\phi_2=\phi_3=0$ for IH
(the largest eigenvalue of the matrix of neutrino Yukawa couplings
is $0.00128 \times \sqrt{m_N}$).
}
\end{figure}

We plot the significance versus the mass of heavy neutrino in Fig. \ref{s}
for NH and IH with parameters given in the caption
and with integrated luminosity $500~{\rm fb}^{-1}$ as an illustration.
In the case of NH, we choose to have $|R_{\mu N}|$
about $10$ times larger than $|R_{eN}|$ for $\delta_{\rm CP}=\pi/2$.
$R_{\tau N}$ is of the same magnitude as $R_{\mu N}$,
so the dominant decay channels are $\mu$ and $\tau$ channel
which dominate the total significance in the figure.
For $\delta_{\rm CP}=-\pi/2$, $|R_{\mu N}|$ is of the same size of $|R_{\tau N}|$,
but approximately 2 times larger than $|R_{eN}|$.
Furthermore, the backgrounds for $\mu$ and $\tau$ channels
are several times smaller than $e$ channel.
So, the $\mu$ and $\tau$ channels are still dominant
in $e+\mu+\tau$ significance.

As can be seen in Fig. \ref{s},  in the case of NH, a heavy sterile neutrino with a mass less than about $152 ~{\rm GeV}$(
$|R_{eN}|\sim 0.0032$, $|R_{\mu N}|\sim |R_{\tau N}| \sim 0.034$)
can be discovered for $\delta_{\rm CP}=\pi/2$.
For $\delta_{\rm CP}=-\pi/2$,  a heavy sterile neutrino with a mass less than around $206 ~{\rm GeV}$
($|R_{eN}|\sim 0.015$, $|R_{\mu N}|\sim |R_{\tau N}| \sim 0.028$) can be discovered. 
As can be seen in the above example, the  case with $\delta_{\rm CP}=-\pi/2$ has a larger $|R_{eN}|$.
This larger value of $|R_{eN}|$ enhances the t-channel production process and gives rise to a larger production rate of heavy sterile neutrino.
Meanwhile, the $\mu$ or $\tau$ channel decay of N is still dominating over the $e$ channel, so
the significance increases a lot from the case of $\delta_{\rm CP}=\pi/2$ to the case of $\delta_{\rm CP}=-\pi/2$.

In the case of IH, we choose to have similar magnitude
of $|R_{eN}|$, $|R_{\mu N}|$ and $|R_{\tau N}|$
for both cases of $\delta_{\rm CP}=\pi/2$ and $\delta_{\rm CP}=-\pi/2$.
All three $e$, $\mu$, and $\tau$ decay channels have
comparable contributions to the total significance.
For Dirac phase of
both cases of $\delta_{\rm CP}=\pi/2$ and $\delta_{\rm CP}=-\pi/2$,
the magnitude of $|R_{eN}|$ has the same size,
and so does $|R_{eN}|^2 + |R_{\mu N}|^2 + |R_{\tau N}|^2$.
This leads to the same production rate of $e^+ e^- \to \nu N$ 
and the same $ljj$  decays of N for both cases of $\delta_{\rm CP}=\pi/2$ and $\delta_{\rm CP}=-\pi/2$.
So, the total $e+\mu+\tau$ significances are the same for both cases of $\delta_{\rm CP}=\pi/2$ and $\delta_{\rm CP}=-\pi/2$.
However, there is a difference between $e+\mu$ significances for these two cases.
One can see in Fig. \ref{s}  that for IH a heavy neutrino with mass less than about $162 {\rm GeV}$
can be discovered. The corresponding mixing parameters in the figure are $|R_{eN}|\sim 0.0086$, $|R_{\mu N}|\sim 0.0072$,
$|R_{\tau N}| \sim 0.0051$ for $\delta_{\rm CP}=\pi/2$,
and $|R_{eN}|\sim 0.0086$, $|R_{\mu N}|\sim 0.0053$,
$|R_{\tau N}| \sim 0.0071$ for $\delta_{\rm CP}=-\pi/2$. 

\begin{figure}[!htb]
\centerline{ \includegraphics[width=8cm]{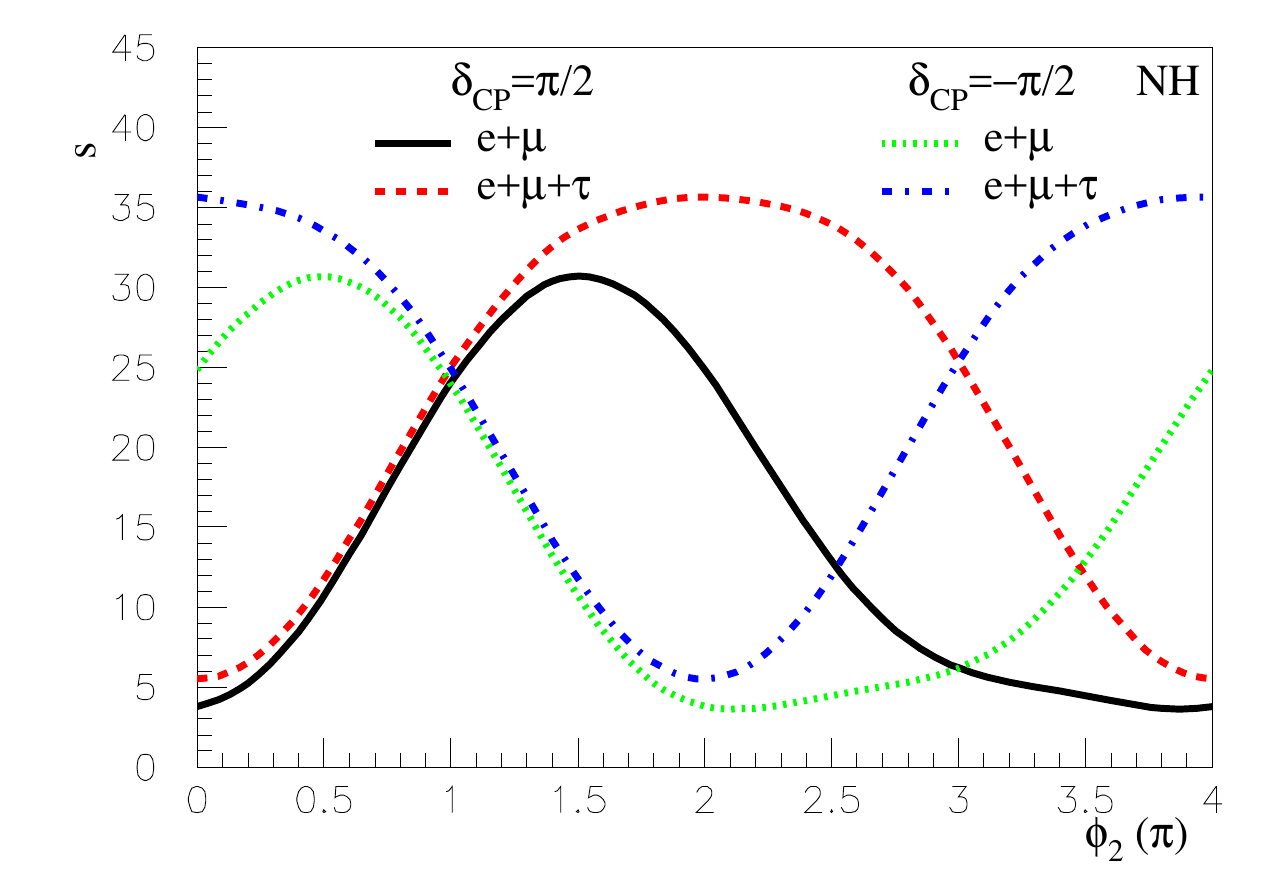}
 \includegraphics[width=8cm]{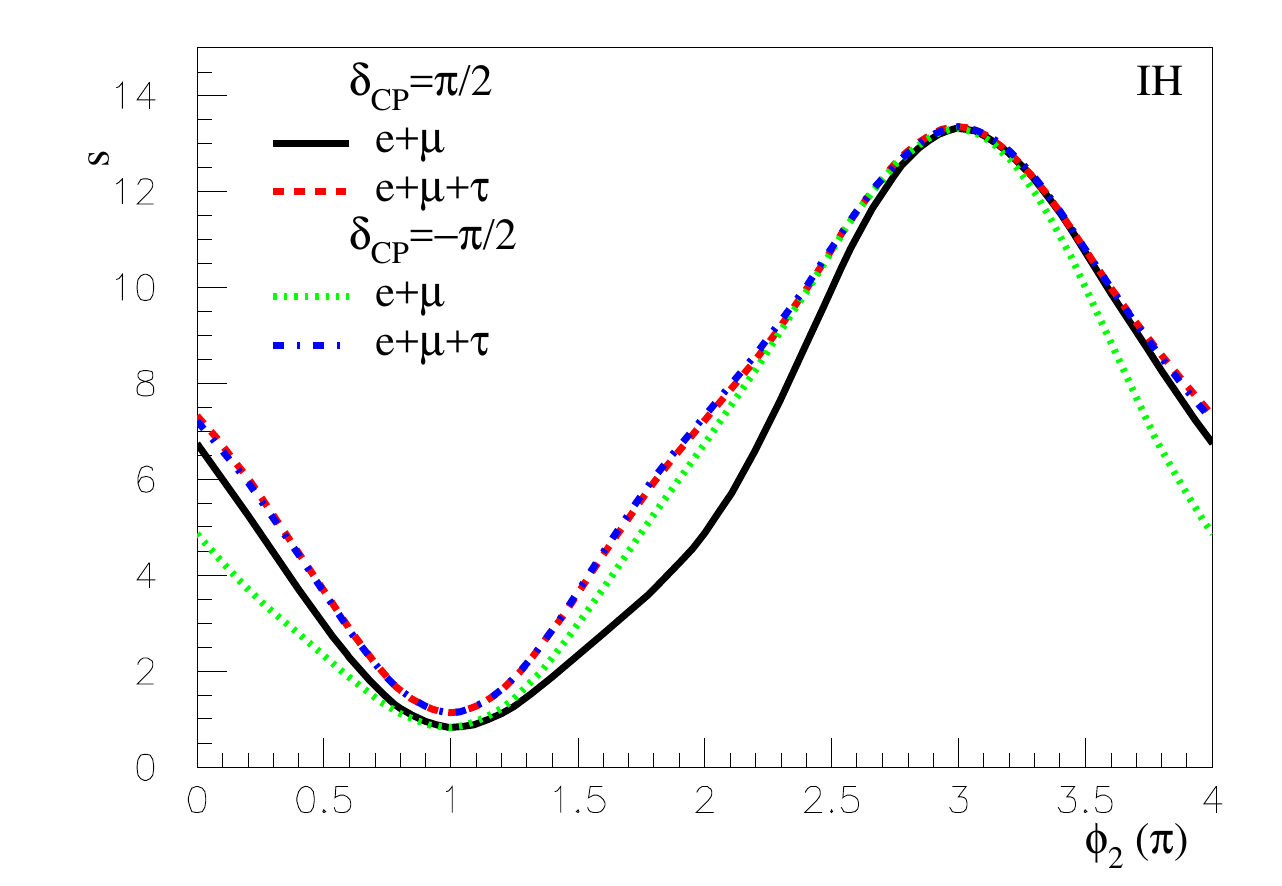}}
\caption{\label{phi}
The significance $s$ vs $\phi_2$ for NH (left) and IH (right)
with integrated luminosity $500{\rm fb}^{-1}$
for a heavy neutrino mass of $150 {\rm GeV}$.
We choose $e^y=5000$, $\delta_{\rm CP}=\pm \pi/2$,
$\phi_1=\phi_3=0$ for NH
(the largest eigenvalue of the matrix of neutrino Yukawa couplings
is $0.0606$),
and $e^y=1000$, $\delta_{\rm CP}=\pm \pi/2$,
$\phi_1=\phi_3=0$ for IH
(the largest eigenvalue of the matrix of neutrino Yukawa couplings
is $0.0156$).
}
\end{figure}

\begin{figure}[!htb]
\centerline{ \includegraphics[width=8cm]{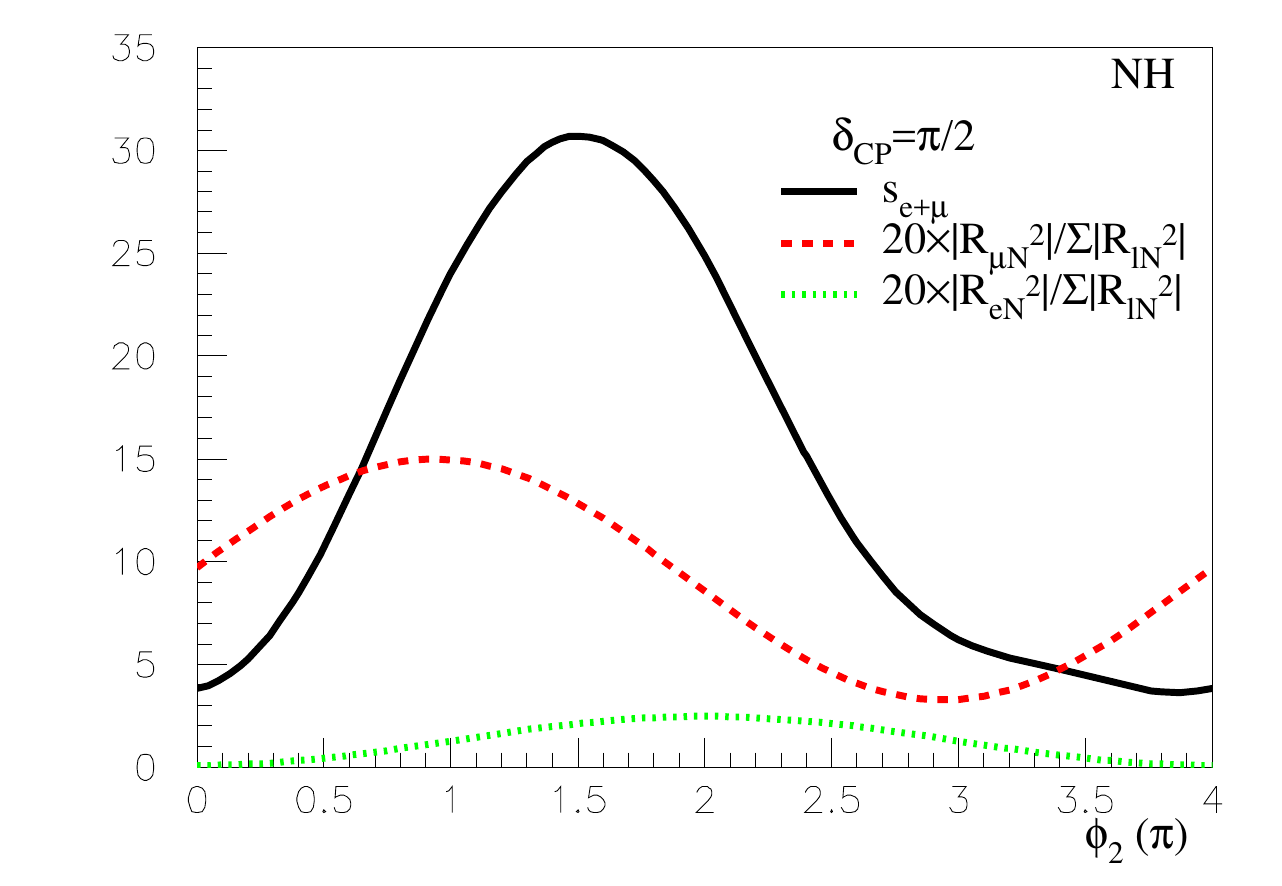} }
\caption{\label{checkphi}
The significance $s$ vs $\phi_2$ for NH with $\delta_{\rm CP}=\pi/2$
with $e^y=5000$, $\phi_1=\phi_3=0$
(the largest eigenvalue of the matrix of neutrino Yukawa couplings
is $0.0606$),
and integrated luminosity $500{\rm fb}^{-1}$.
}
\end{figure}

In Fig. \ref{phi}, we also plot the total significance as a function of the Majorana phase $\phi_2$  for a heavy neutrino mass of $150 ~{\rm GeV}$
and integrated luminosity of $500 ~{\rm fb}^{-1}$ for both NH and IH.
The significance depends on both Dirac phase of $\delta_{\rm CP}$
and Majorana phase $\phi_2$.
In the case of NH with $\delta_{\rm CP}=\pi/2$,
there is a bump at $\phi_2 \sim 1.5 \pi$ for $e+\mu$ significance. On the other hand,
the bump is at around $2\pi$ for $e+\mu+\tau$ significance. 
This is because $|R_{e N}|^2/\sum |R_{lN}|^2$ increases as $\phi_2$ increases from $0$
to $2 \pi$, as can be seen in Fig. \ref{checkphi}.  Since $\sum |R_{lN}|^2$ is a constant
when varying $\phi_2$, as can be easily checked using (\ref{mixing1}) and (\ref{mixing2}),
$|R_{e N}|^2$ increases as $\phi_2$ increases from $0$ to $2 \pi$ and peaks at $\phi_2=2\pi$.
Consequently, the t-channel production process, the dominating production process, 
increases as $\phi_2$ increases from $0$ to $2 \pi$.
This is why the plot of $e+\mu+\tau$ significance peaks at $\phi_2=2\pi$ in the case of NH with $\delta_{\rm CP}=\pi/2$.
For $e+\mu$ significance, it is dominated by the $\mu jj$ events, as explained before.
As $\phi_2$ increases, $|R_{\mu N}|^2/\sum |R_{lN}|^2$ peaks
at $\phi_2 \sim \pi$.  For $\phi_2$ larger than around $\pi$, 
the branching fraction of $N \to \mu jj$ decay starts to decrease,
which is compensated by
the increase of the production cross section of $e^+e^-\to N\nu$.
Then, the signature of $\mu jj$ events will increase first
and then decrease as $\phi_2$ increases from $\pi$ to $2\pi$.  
This makes $e+\mu$ significance having a peak at a position less than $2\pi$, as can seen in Fig. \ref{phi}.
Variation of significance in other cases can be similarly understood.

\begin{figure}[!htb]
\centerline{ \includegraphics[width=8cm]{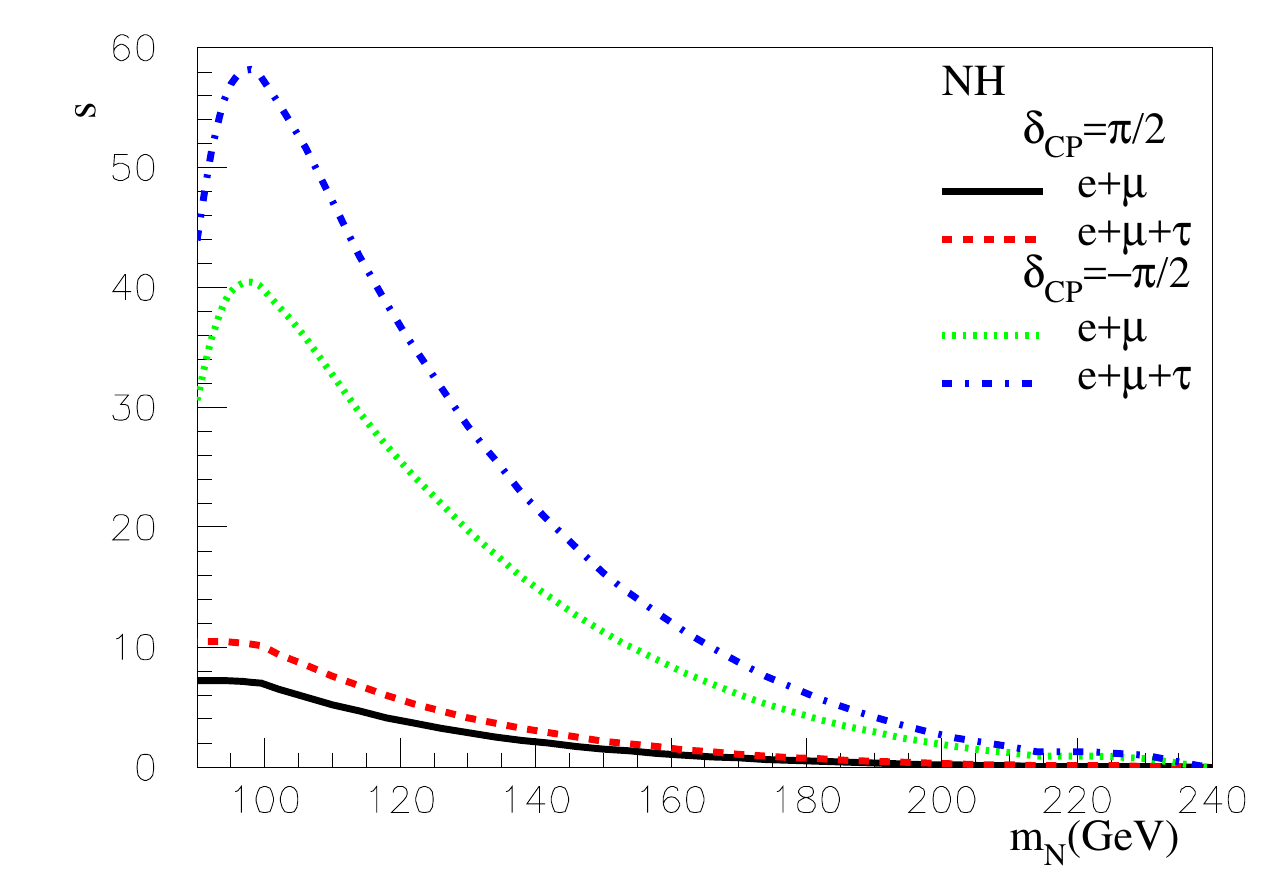}
 \includegraphics[width=8cm]{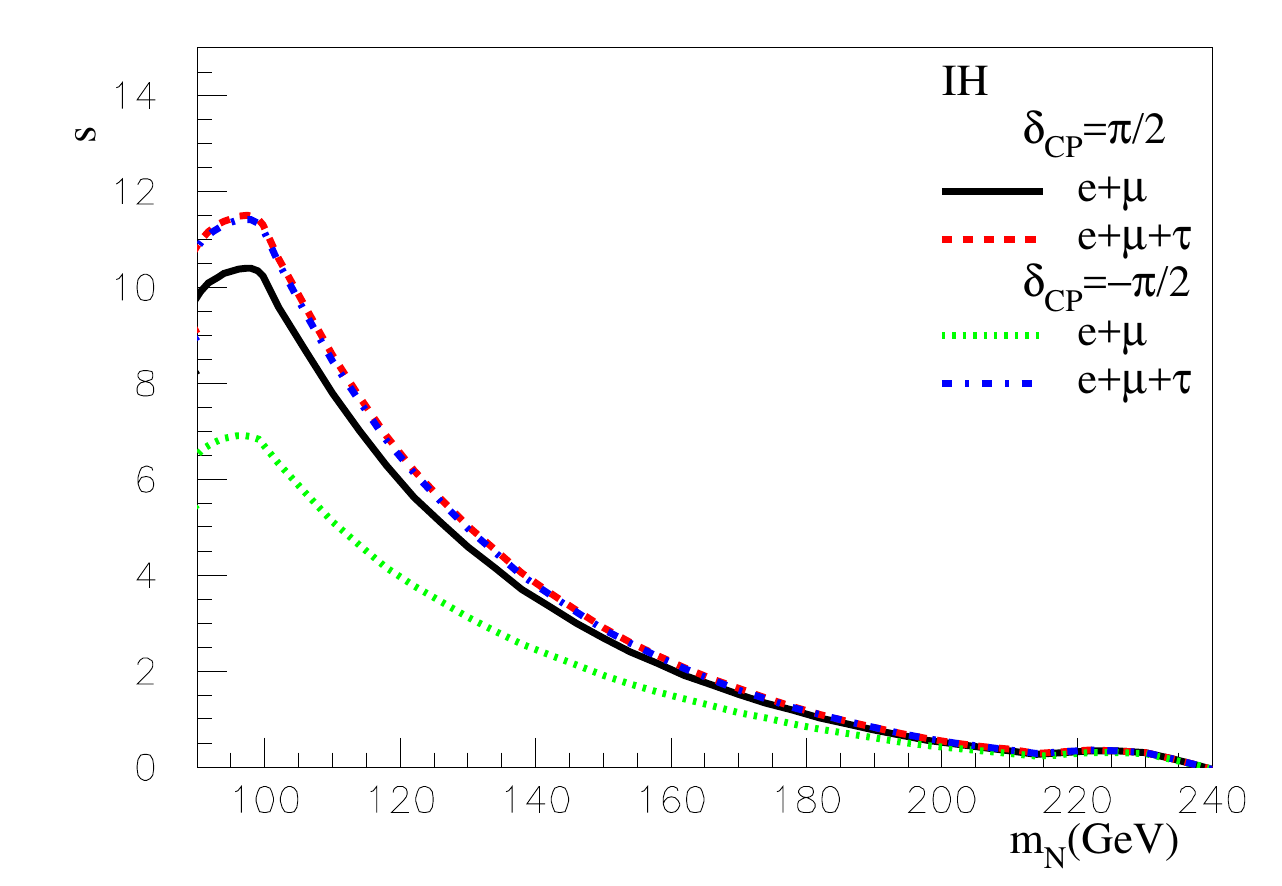}}
\caption{\label{s5ab}
The significance $s$ vs $m_N$ for NH (left) and IH (right)
with integrated luminosity $5{\rm ab}^{-1}$.
We choose $e^y=1750$, $\delta_{\rm CP}=\pm \pi/2$,
$\phi_1=\phi_2=\phi_3=0$ for NH
(the largest eigenvalue of the matrix of neutrino Yukawa couplings
is $0.00173\times \sqrt{m_N}$),
and $e^y=350$, $\delta_{\rm CP}=\pm \pi/2$,
$\phi_1=\phi_2=\phi_3=0$ for IH
(the largest eigenvalue of the matrix of neutrino Yukawa couplings
is $0.000447 \times \sqrt{m_N}$).
}
\end{figure}

In Fig.~\ref{s5ab}, we present the significance as a function of heavy neutrino mass
with integrated lumilosity $5 {\rm ab}^{-1}$.
In the case of NH, a heavy neutrino mass less than about
$124 {\rm GeV}$
($|R_{eN}|\sim 0.0012$, $|R_{\mu N}|\sim |R_{\tau N}| \sim 0.013$)
for $\delta_{\rm CP}=\pi/2$,
and $184 {\rm GeV}$
($|R_{eN}|\sim 0.0055$, $|R_{\mu N}|\sim |R_{\tau N}| \sim 0.010$)
for $\delta_{\rm CP}=-\pi/2$
can be discovered at CEPC.
In the case of IH, a heavy neutrino mass less than about
$130 {\rm GeV}$ can be discovered.
The corresponding mixing parameters are
$|R_{eN}|\sim 0.0034$, $|R_{\mu N}|\sim 0.0028$, $|R_{\tau N}| \sim 0.0020$
for $\delta_{\rm CP}=\pi/2$,
and $|R_{eN}|\sim 0.0034$, $|R_{\mu N}|\sim 0.0021$, $|R_{\tau N}| \sim 0.0028$
for $\delta_{\rm CP}=-\pi/2$.

\begin{figure}[!htb]
\includegraphics[width=8cm]{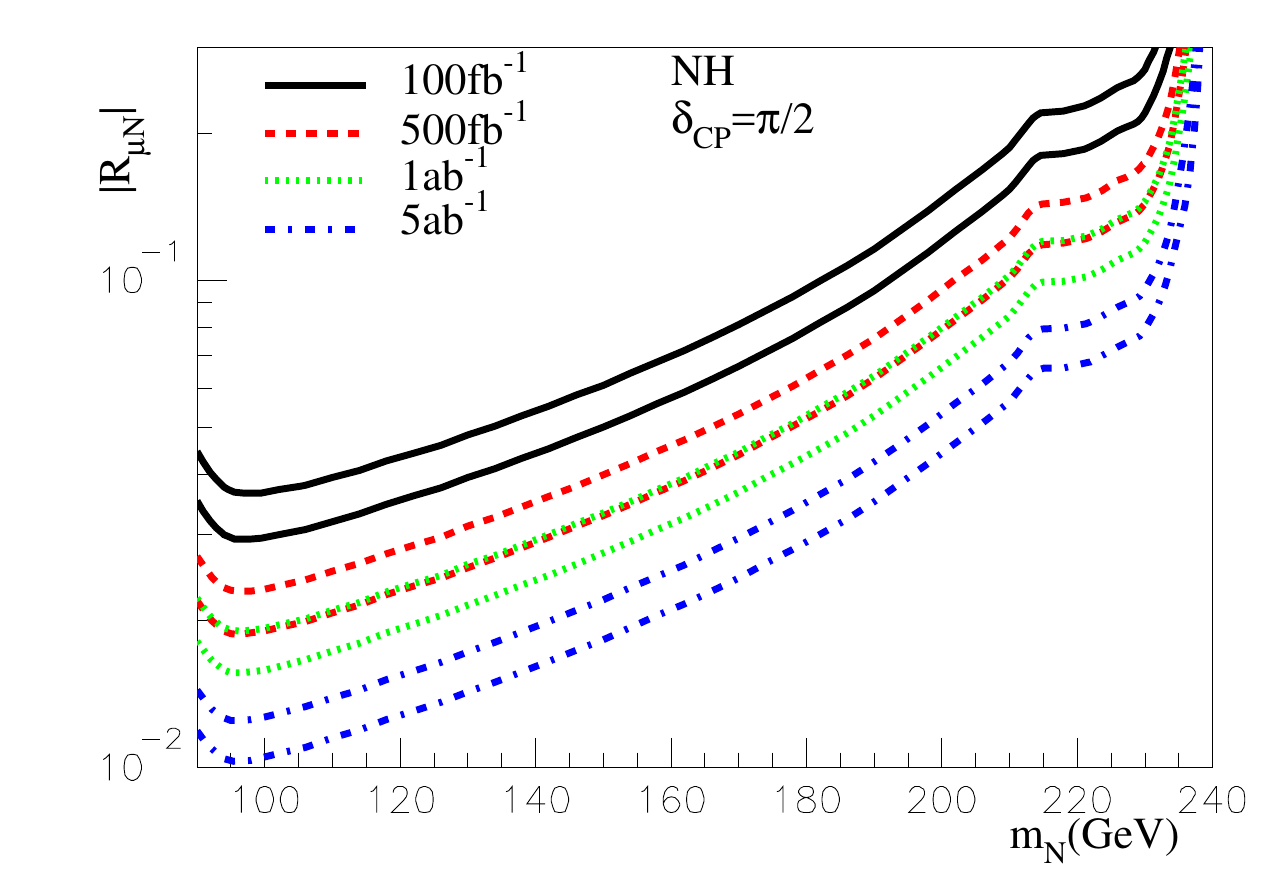}
\includegraphics[width=8cm]{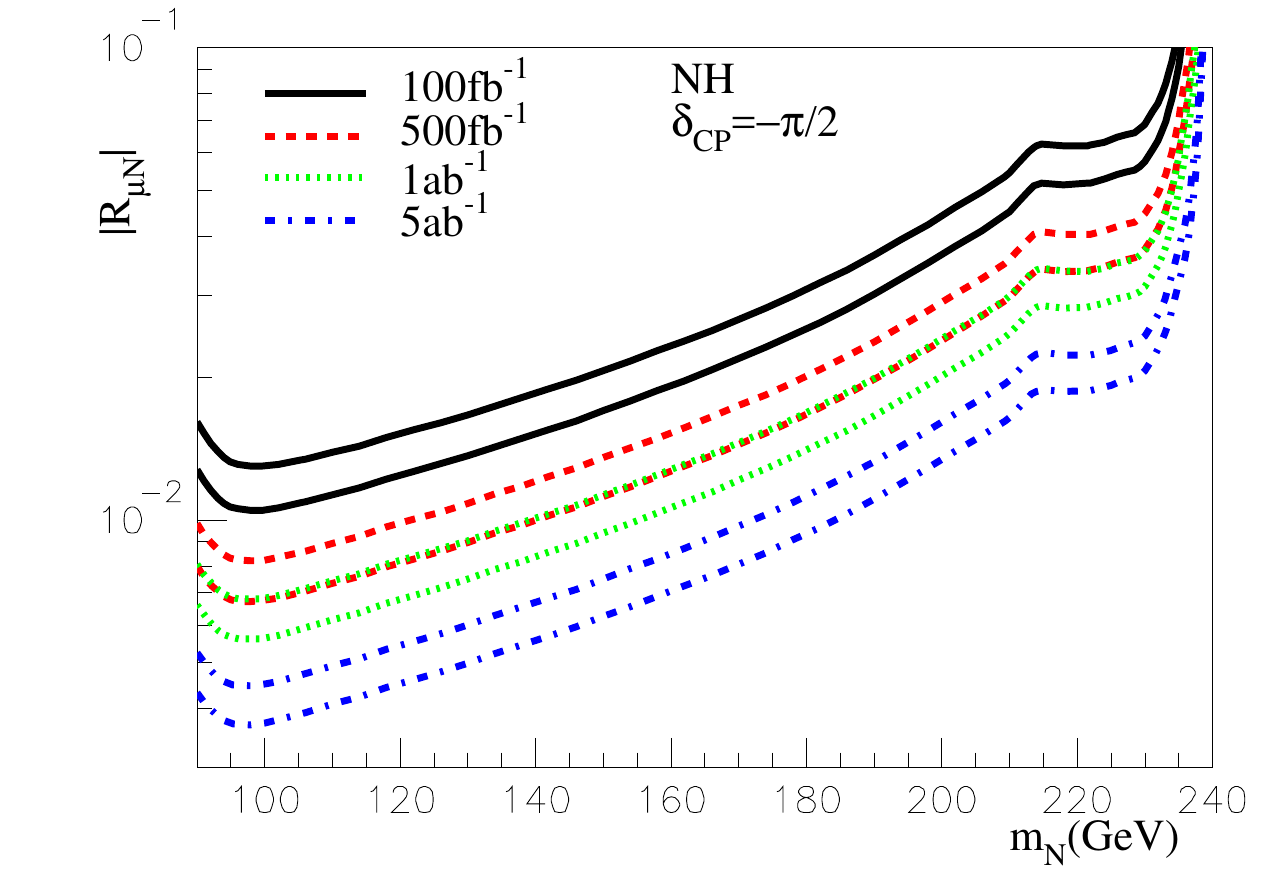} \\
\includegraphics[width=8cm]{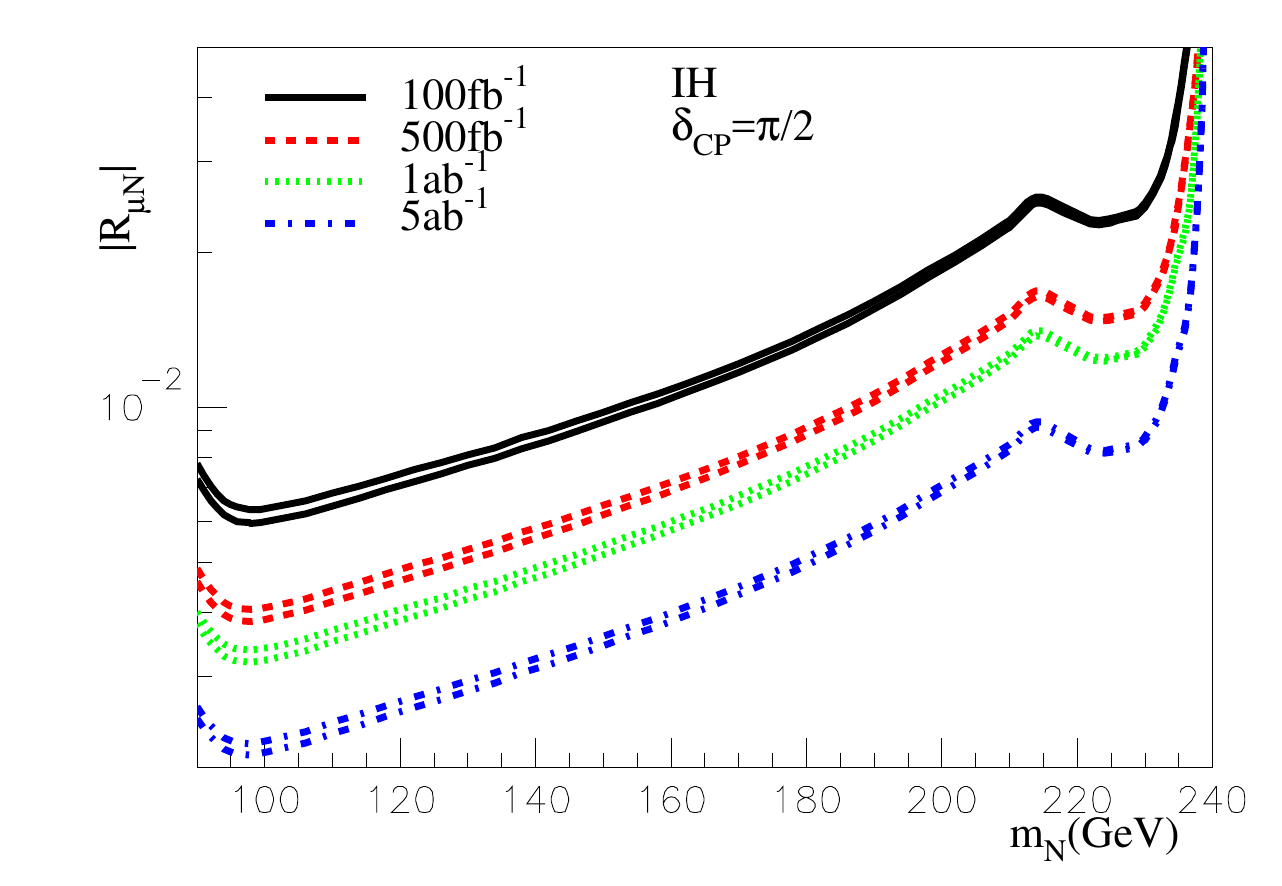}
\includegraphics[width=8cm]{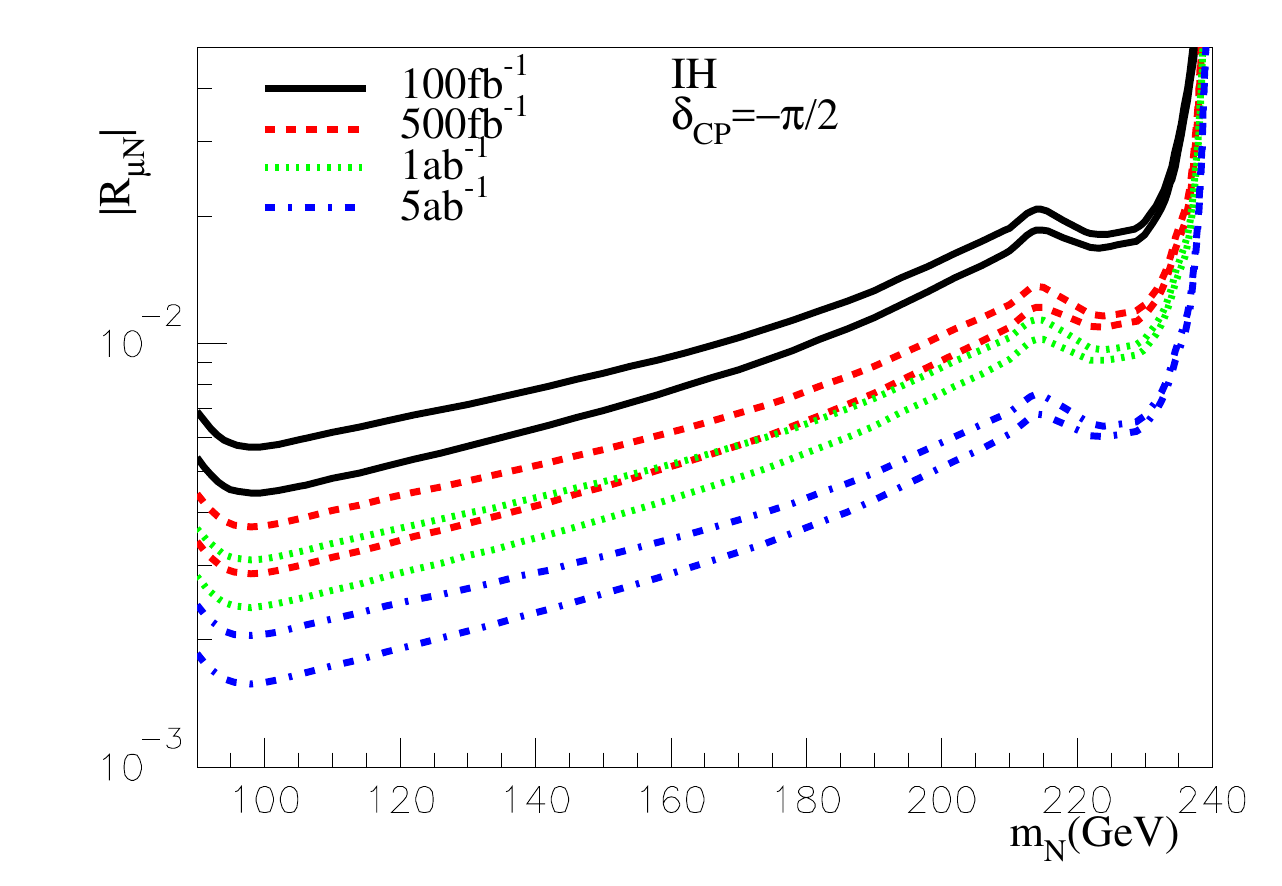}
\caption{\label{RlNcorrelation}
Sensitivity to $|R_{\mu N}|$ with significance $s=5$
for the cases of NH and IH and Dirac phase $\delta_{\rm CP}=\pm \pi/2$,
$\phi_1=\phi_2=\phi_3=0$.
The curves of different type in each plot from up to down correspond to
integrated luminosities $100{\rm fb}^{-1}$, $500{\rm fb}^{-1}$,
$1{\rm ab}^{-1}$ and $5{\rm ab}^{-1}$.
The upper (lower) curve of the same type corresponds
to $e+\mu$ ($e+\mu+\tau$) significances in each plot, respectively.
}
\end{figure}

In Fig.~\ref{RlNcorrelation}, we plot the potential of
probing $|R_{\mu N}|$ for $e+\mu$ (or $e+\mu+\tau$) significance $s=5$
with the integrated luminosities $5{\rm ab}^{-1}$, $1{\rm ab}^{-1}$,
$500{\rm fb}^{-1}$ and $100{\rm fb}^{-1}$ at CEPC
for different cases of NH and IH, and Dirac phase $\delta_{\rm CP}=\pm \pi/2$.
For each case, the ratio of $|R_{eN}| : |R_{\mu N}| : |R_{\tau N}|$
is fixed for given Dirac phase and Majorana phases,
therefore we only plot $|R_{\mu N}|$ for illustration.
In the case of NH, the ratio of $|R_{eN}| : |R_{\mu N}| : |R_{\tau N}|$
is $0.0945 : 1 : 1.03$ ($0.537 : 1 : 1.02$)
for $\delta_{\rm CP}=\pi/2$ ($\delta_{\rm CP}=-\pi/2$).
In the case of IH, the ratio of $|R_{eN}| : |R_{\mu N}| : |R_{\tau N}|$
is $1.19 : 1 : 0.709$ ($1.61 : 1 : 1.32$)
for $\delta_{\rm CP}=\pi/2$ ($\delta_{\rm CP}=-\pi/2$).
In the case of NH with $\delta_{\rm CP}=-\pi/2$ and
the case of IH with $\delta_{\rm CP}=\pm \pi/2$,
the three $|R_{lN}|$ are of similar magnitude,
so $|R_{\mu N}|$ can be probed to order of $10^{-3}$ for $5{\rm ab}^{-1}$
with enhanced production rate due to large $|R_{eN}|$.
In the case of NH with $\delta_{\rm CP}=\pi/2$,
$|R_{eN}|$ is almost $10$ times smaller than $|R_{\mu N}|$,
then the mixing $|R_{\mu N}|$ of order of $10^{-2}$ can be probed
for $5{\rm ab}^{-1}$, which is similar to the case with a single
non-zero $|R_{\mu N}|$ as given in Fig. 6.

To conclude, in the low energy see-saw model,
due to the correlation of three different $R_{lN}$,
sizable $|R_{e N}|$ leads to t-channel production of heavy sterile neutrino and
can give rise to a quite large total production cross section of  $e^+e^-\to N\nu$ prcocess.
The $N \to ljj$ events, on the other hand, can be dominated by $\mu jj$ and $\tau jj$ events
because $|R_{\mu N}|^2+|R_{\tau N}|^2$ can be much larger than $|R_{eN}|^2$ as can be seen in Fig. \ref{mixing-relation}.  
For NH, in particular, $|R_{\mu N}|^2+|R_{\tau N}|^2$ is always much larger than $|R_{eN}|^2$.
In this case, $e+\mu$ significance  and $e+\mu+\tau$ significance
can be quite different, as can be seen in the left panel of Fig. \ref{phi}. 
On the other hand, for IH, $|R_{\mu N}|^2+|R_{\tau N}|^2$ can be of similar size of $|R_{eN}|^2$ and even much smaller than
$|R_{eN}|^2$. In this case, $e+\mu$ significance  and $e+\mu+\tau$ significance would not be very different.
This is the case for the right panel in Fig. \ref{phi}.  So analyzing the dominating the $N$ decay channel 
and the difference of $e+\mu$ and $e+\mu+\tau$ significances can give hints on
the mass hierarchy of neutrinos. In particular, if the dominating $N\to ljj$ events are $ejj$ events, it has to be
IH.

\bigskip
\section{Conclusion}
In summary, we have studied the production, decay and signature in $ljj$ events
of heavy Majorana-type sterile neutrino of mass around 100 GeV
 at future CEPC.  We study carefully the tree-level decay of heavy sterile neutrinos by carefully taking into account
the propagator of bosons, such as $W$ and $Z$.  Effects of on-shell and off-shell
$W$ and $Z$ bosons are all taken into account by including the width of W and Z in the propagators.
We obtain analytic formula for tree-level decay of heavy sterile neutrinos which is valid for mass from tens of GeV to
hundreds GeV and higher energy. The formula are valid in particular for mass $m_N$ around the masses of bosons.

For convenience and for later discussion in low energy see-saw model of heavy sterile neutrino,
we have first studied the production of a single heavy sterile neutrino at CEPC and its signature. 
Although the mixing of a single heavy sterile neutrino with active neutrino is strongly constrained
by the $0\nu\beta\beta$ experiment, 
the study of the signature of a single heavy sterile neutrino is also of interests for itself,
since some other particles or mechanism, e.g. extra scalars or type-II seesaw, may exist to ease the constraint. 
We have shown that for a single heavy sterile neutrino, an electron positron collider such as CEPC
is more sensitive to the mixing of heavy sterile neutrino with electron (anti)neutrino, than the
mixing with muon or tau (anti)neutrino. For the former, the production of $N$ is associated with
the production of an electron neutrino or anti-neutrino and can go through t-channel. The cross section
of the t-channel process can be two orders of magnitude larger than the cross section of the
s-channel process which is responsible for probing the magnitude of the mixing with muon or tau (anti)neutrino.
We found that for  an integrated luminosity 5 ab$^{-1}$, CEPC can reach a $5 \sigma$ sensitivity of $R_{eN}$, the mixing
of the sterile neutrino with active neutrino, to a value as small as $|R_{eN}| =10^{-3}$.
For the mixing with muon and tau (anti)neutrino $R_{\mu N}$ and $R_{\tau N}$, the $5 \sigma$ sensitivity
can reach $|R_{\mu N,\tau N}|\approx 10^{-2}$.

We also study the production of heavy sterile neutrinos in a low energy see-saw model and their signature at CEPC.
In this model, two heavy sterile neutrinos exist so that an explanation of the masses and mixings of active neutrinos
is available using see-saw mechanism. In this model, the mixings of these two heavy sterile neutrinos with active neutrinos, 
$R_{lN_1}$ and $R_{lN_2}$, are forced to have the same magnitude for all $l$,  if we want these mixings to be large. 
In this case, the masses of these two sterile neutrinos are found to be degenerate or quasi-degenerate if considering into account
the constraint from $0\nu \beta \beta$ experiment.

So the signature of these two heavy sterile neutrinos 
are just the double of the signature of
a single heavy sterile neutrino discussed above. 
The major difference
compared with the case of a single heavy sterile neutrino is that
the mixing $R_{lN_1}$ is no longer arbitrary for
different $l$. Instead, values of $R_{l N_1}$ for different $l$
have some correlations. We take these facts into account.
We find that the Dirac CP phase $\delta_{\rm CP}$
in the PMNS mixing matrix of active neutrinos and
Majorana phases affect the mixing $R_{lN}$,
and change the relative significance of $ejj$, $\mu jj$ and $\tau jj$ events. 
So a search for all $3$ lepton channels are helpful
to constrain the model.
With sizable $R_{eN}$, the significance of
both $\mu$ and $\tau$ channel will be enhanced,
and further constrain $R_{\mu N}$ and $R_{\tau N}$
compared with the case with only a single mixing.

We further note that although our analysis is for CEPC running at 240 GeV,
it can also be applied to ILC running at around 250 GeV without much modification~\cite{ILC250}.

\bigskip
\section*{Acknowledgments}
This research is supported in part by the Natural Science Foundation of
China(NSFC), Grant No. 11135009 and No. 11375065, and in part by the
Shanghai Key Laboratory of Particle Physics and Cosmology,
Grant No. 15DZ2272100.
XHW would like to thank Qi-Shu Yan for helpful discussions
on MadGraph and CEPC.

\bigskip
\section{Appendix}
In this section we summarize the tree level decay rate of sterile neutrino decaying to three final fermions through
interaction with Z and W bosons induced by mixing with active neutrinos. Effects of on-shell and off-shell
Z and W bosons are all taken into account by including the width of W and Z in the propagators.
For example, for $N \to l_1^- l_2^+ \nu_{l_2}$ and $l_1\neq l_2$,  the decay rate is obtained as follows
\begin{eqnarray}
\Gamma(N \to l_1^- l_2^+ \nu_{l_2})=|R_{l_1 N}|^2\frac{G_F^2 m_N}{\pi^3} 
\int^{m_N \over 2}_0 dE_1\int^{m_N \over 2}_{{m_N \over 2}-E_1} dE_2 ~|X_W|^2\frac{1}{2}(m_N-2 E_2)E_2, \label{decayR0}
\end{eqnarray}
where $X_W$ comes from the propagator of W boson and is
\begin{eqnarray}
X_W=\frac{m_W^2}{q^2-m_W^2+i\Gamma_W m_W},\label{XW}
\end{eqnarray}
where $q^2=m^2_N-2 m_N E_1$ and $\Gamma_W$ is the total decay rate of $W$. 
$q=p-p_1$ is the four momentum of the W boson where
$p$ and $p_1$ are the four momenta of $N$ and $l_1$ respectively. So $q^2=m^2_N-2 m_N E_1$ when considering
the decay of $N$ at rest and neglecting the mass of $l_1$ with $E_1$ the energy of $l_1$.  After performing integration 
in (\ref{decayR0}) we can get a formula for the decay rate as a function of $m_N$, $m_W$ and $\Gamma_W$.
Similarly we can get formula for other decays through Z boson exchange.

In the following we summarize the results\\
1)For $N \to l_1^- l_2^+ \nu_{l_2}$, $N \to l_1^+ l_2^- {\bar \nu}_{l_2}$ and $l_1\neq l_2$
\begin{eqnarray}
\Gamma(N \to l_1^- l_2^+ \nu_{l_2})=\Gamma(N \to l_1^+ l_2^- {\bar \nu}_{l_2})
=|R_{l_1 N}|^2\frac{G_F^2 m_N^5}{\pi^3} F_N(m_N,m_W,\Gamma_W), \label{decayR1}
\end{eqnarray}
where $F_N$ is a dimensionless function and is given in (\ref{FN}) below. 

2) For $N \to l^- q_1 {\bar q}_2$, $N \to l^+ {\bar q}_1 q_2$
\begin{eqnarray}
\Gamma(N\to  l^- q_1 {\bar q}_2)=\Gamma(N\to l^+ {\bar q}_1 q_2)
=|R_{l N}|^2\frac{G_F^2 m_N^5}{\pi^3} N_C F_N(m_N,m_W,\Gamma_W) |K_{q_1 q_2}|^2. \label{decayR2}
\end{eqnarray}
$K_{q_1 q_2}$ is the CKM matrix element in $(q_1,q_2)$ entry, $N_C=3$ the number of color degrees of freedom of quarks. 

3) For $N\to l^- l^+ \nu_l$, $N\to l^+l^- {\bar \nu}_l$
\begin{eqnarray}
&&\Gamma(N\to l^- l^+ \nu_l)=\Gamma(N\to l^+ l^- {\bar \nu}_l) \nonumber \\
&& =|R_{l N}|^2\frac{G_F^2 m_N^5}{\pi^3} [ F_N(m_N,m_W,\Gamma_W)+(C_L^2+C_R^2)F_N(m_N,m_Z,\Gamma_Z) \nonumber \\
&&+2 C_L  ~F_S(m_N, m_W,\Gamma_W,m_Z,\Gamma_Z)], \label{decayR3}
\end{eqnarray}
where $C_{L,R}$ is given in (\ref{couplings0-1}), $F_S$ is a dimensionless function and is given below in (\ref{FS}). 

4) For $N\to \nu_l {\bar l}' l'$ and $N\to {\bar \nu}_l l' {\bar l}'$
\begin{eqnarray}
\Gamma(N\to \nu_l {\bar l}' l')=\Gamma(N\to {\bar \nu}_l l' {\bar l}')=
|R_{l N}|^2\frac{G_F^2 m_N^5}{\pi^3} (C_L^2+C_R^2) F_N(m_N,m_Z,\Gamma_Z). \label{decayR4}
\end{eqnarray}

5) For $N \to \nu_l q {\bar q}$ and $N \to {\bar \nu}_l {\bar q} q$
\begin{eqnarray}
\Gamma(N\to \nu_l {\bar l}' l')=\Gamma(N\to {\bar \nu}_l l' {\bar l}')=
|R_{l N}|^2\frac{G_F^2 m_N^5}{\pi^3}N_C [(C^q_L)^2+(C^q_R)^2] F_N(m_N,m_Z,\Gamma_Z), \label{decayR5}
\end{eqnarray}
where $q=u,d,c,s,b$ for $m_N < 2 m_t$ and $C^q_{L,R}$ is given in (\ref{couplings0-2}) and (\ref{couplings0-3}).

6)For $N \to \nu_l {\nu}_{l'} {\bar \nu}_{l'}$ and $N \to {\bar \nu}_l {\bar \nu}_{l'}  {\nu}_{l'} $, $l\neq l'$
\begin{eqnarray}
\Gamma(N \to \nu_l {\nu}_{l'} {\bar \nu}_{l'})=\Gamma(N \to {\bar \nu}_l {\nu}_{l'} {\bar \nu}_{l'})
=|R_{l N}|^2\frac{G_F^2 m_N^5}{\pi^3} C_\nu^2 F_N(m_N,m_Z,\Gamma_Z), \label{decayR6}
\end{eqnarray}
where $C_\nu =1/2$.

7) For $N \to \nu_l {\nu}_l {\bar \nu}_l$ and $N \to {\bar \nu}_l {\bar \nu}_l {\nu}_l $
\begin{eqnarray}
\Gamma(N \to \nu_l {\nu}_{l} {\bar \nu}_{l})=\Gamma(N \to {\bar \nu}_l {\nu}_{l} {\bar \nu}_{l})
=|R_{l N}|^2\frac{G_F^2 m_N^5}{\pi^3} 4 C_\nu^2 F_N(m_N,m_Z,\Gamma_Z). \label{decayR7}
\end{eqnarray}

Couplings $C_L$, $C_R$ etc. which appear in expressions above, are given as
\begin{eqnarray}
& C_L=-\frac{1}{2}+\sin^2\theta_W,~C_R=\sin^2\theta_W,\label{couplings0-1} \\
& C_L^u=\frac{1}{2}-\frac{2}{3} \sin^2\theta_W, C_R^u=-\frac{2}{3} \sin^2\theta_W, \label{couplings0-2} \\
& C_L^u=-\frac{1}{2}+\frac{1}{3} \sin^2\theta_W, C_R^u=\frac{1}{3} \sin^2\theta_W. \label{couplings0-3}
\end{eqnarray}

For mass $m_N$, $m_X$ and decay rate $\Gamma_X$, the function $F_N$ used above is
\begin{eqnarray}
&& F_N(m_N, m_X,\Gamma_X)=  \frac{m_X^4}{96 m_N^8} \bigg \{  -2 m_N^2 (m_N^2-m_X^2) \nonumber \\
&& +(A_X+C_X \Gamma_X^2 m_X^2)\frac{1}{\Gamma_X m_X} \bigg[  arctan\bigg(  \frac{m_N^2-m_X^2}{\Gamma_X m_X} \bigg) 
-arctan\bigg(\frac{-m_X^2}{\Gamma_X m_X} \bigg) \bigg] \nonumber \\
&& -\frac{1}{2}(B_X+2\Gamma_X^2 m_X^2) ln\bigg ( \frac{\Gamma_X^2 m_X^2+(m_N^2-m_X^2)^2}{\Gamma_X^2m_X^2+m_X^4} \bigg)
\bigg\}  \label{FN}
\end{eqnarray}
where 
\begin{eqnarray}
A_X= (m_N^2-m_X^2)^2 (m_N^2+2 m_X^2), ~B_X=6(m_N^2-m_X^2)m_X^2,~C_X=3(m_N^2-2 m_X^2)
\end{eqnarray}
Function $F_S$ in (\ref{decayR3}) is given as
\begin{eqnarray}
F_S=\frac{1}{m_N^4}\int^{m_N \over 2}_0 dE_1 \int^{m_N \over 2}_{{m_N \over 2}-E_1}dE_2(X_W X_Z^* +X_W^* X_Z) 
\frac{1}{2}(m_N-2 E_2) E_2, \label{FS}
\end{eqnarray}
where
\begin{eqnarray}
X_Z=\frac{m_Z^2}{q^2_3-m_Z^2+i\Gamma_Z m_Z}.\label{XZ}
\end{eqnarray}
$q_3^2=m_N^2-2 m_N E_3$ with $E_3=m_N-E_1-E_2$ when considering
the decay of $N$ at rest and neglecting the mass of final fermions.
(\ref{FS}) can not be obtained as an explicit function of $m_N$, $m_W$ and $m_Z$. We
can calculate this function numerically.

For $N \to \nu H$ decay, the effect described here can be similarly obtained with the introduction of a function
$F_N(m_N,m_H,\Gamma_H)$.  For example, for $N\to \nu_l {\bar f} f$ and $N \to {\bar \nu}_l {\bar f} f$
\begin{eqnarray}
\Gamma(N\to  \nu_l {\bar f} f)=\Gamma(N \to {\bar \nu}_l {\bar f} f)
=\frac{g^2 m_N^7 |R_{lN}|^2 y_f^2}{16\pi^3 m_W^2 m_H^4} N_f F_N(m_N,m_H,\Gamma_H) , \label{decayR-NtoH}
\end{eqnarray}
where $y_f$ is the Yukawa coupling of fermion $f$, $N_f=1$ for f  being a lepton and $N_f=3$ for f being a quark.  
Interference of $N$ decay through Z boson and H boson vanishes.
Since the Yukawa coupling to fermion $f$ is always small for $f=b,c,s,d,u$ and leptons, inclusion of $N$ decay through 
the neutral Higgs boson does not change significantly the signature of sterile neutrino $N$ discussed in this article,
as long as we are not going to concentrate on the signature of $N$ coming from $N\to {\nu}_l b{\bar b}$ 
and $N\to {\bar \nu}_l b{\bar b}$ decay. 
  
 In low energy limit $m_N^2 \ll m^2_W$,  we have $|X_W|\approx |X_Z|\approx 1$, the above equations of
 decay rate, (\ref{decayR1}), (\ref{decayR2}), (\ref{decayR3}), (\ref{decayR4}), (\ref{decayR5}), (\ref{decayR6}), (\ref{decayR7}),
 can be simplified to be as follows.\\
1)For $N \to l_1^- l_2^+ \nu_{l_2}$, $N \to l_1^+ l_2^- {\bar \nu}_{l_2}$ and $l_1\neq l_2$
\begin{eqnarray}
\Gamma(N \to l_1^- l_2^+ \nu_{l_2})=\Gamma(N \to l_1^+ l_2^- {\bar \nu}_{l_2})
=|R_{l_1 N}|^2\frac{G_F^2 m_N^5}{192\pi^3}, \label{decayR1-1}
\end{eqnarray}
2) For $N \to l^- q_1 {\bar q}_2$, $N \to l^+ {\bar q}_1 q_2$
\begin{eqnarray}
\Gamma(N\to  l^- q_1 {\bar q}_2)=\Gamma(N\to l^+ {\bar q}_1 q_2)
=|R_{l N}|^2\frac{G_F^2 m_N^5}{192 \pi^3} N_C |K_{q_1 q_2}|^2. \label{decayR1-2}
\end{eqnarray}
3) For $N\to l^- l^+ \nu_l$, $N\to l^+l^- {\bar \nu}_l$
\begin{eqnarray}
\Gamma(N\to l^- l^+ \nu_l)=\Gamma(N\to l^+ l^- {\bar \nu}_l) 
=|R_{l N}|^2\frac{G_F^2 m_N^5}{192\pi^3} [ (1+C_L)^2+C_R^2], \label{decayR1-3}
\end{eqnarray}
4) For $N\to \nu_l {\bar l}' l'$ and $N\to {\bar \nu}_l l' {\bar l}'$
\begin{eqnarray}
\Gamma(N\to \nu_l {\bar l}' l')=\Gamma(N\to {\bar \nu}_l l' {\bar l}')=
|R_{l N}|^2\frac{G_F^2 m_N^5}{192\pi^3} (C_L^2+C_R^2) . \label{decayR1-4}
\end{eqnarray}
5) For $N \to \nu_l q {\bar q}$ and $N \to {\bar \nu}_l {\bar q} q$
\begin{eqnarray}
\Gamma(N\to \nu_l {\bar l}' l')=\Gamma(N\to {\bar \nu}_l l' {\bar l}')=
|R_{l N}|^2\frac{G_F^2 m_N^5}{192 \pi^3}N_C[(C^q_L)^2+(C^q_R)^2] .\label{decayR1-5}
\end{eqnarray}
6)For $N \to \nu_l {\nu}_{l'} {\bar \nu}_{l'}$ and $N \to {\bar \nu}_l {\bar \nu}_{l'}  {\nu}_{l'} $, $l\neq l'$
\begin{eqnarray}
\Gamma(N \to \nu_l {\nu}_{l'} {\bar \nu}_{l'})=\Gamma(N \to {\bar \nu}_l {\nu}_{l'} {\bar \nu}_{l'})
=|R_{l N}|^2\frac{G_F^2 m_N^5}{192\pi^3} C_\nu^2 , \label{decayR1-6}
\end{eqnarray}
7) For $N \to \nu_l {\nu}_l {\bar \nu}_l$ and $N \to {\bar \nu}_l {\bar \nu}_l {\nu}_l $
\begin{eqnarray}
\Gamma(N \to \nu_l {\nu}_{l} {\bar \nu}_{l})=\Gamma(N \to {\bar \nu}_l {\nu}_{l} {\bar \nu}_{l})
=|R_{l N}|^2\frac{G_F^2 m_N^5}{192\pi^3} 4 C_\nu^2 . \label{decayR1-7}
\end{eqnarray}
In all these results, the masses of the final fermions have all been neglected.

\end{document}